\font \eightbf         = cmbx8  
\font \eighti          = cmmi8  
\font \eightit         = cmti8  
\font \eightrm         = cmr8   
\font \eightsl         = cmsl8  
\font \eightsy         = cmsy8  
\font \eighttt         = cmtt8  
\font \tenbf           = cmbx9
\font \teni            = cmmi9
\font \tenit           = cmti9
\font \tenrm           = cmr9
\font \tensl           = cmsl9
\font \tensy           = cmsy9
\font \tentt           = cmtt9

\font \kleinhalbcurs   = cmmib10 at 8pt

\font \sixbf           = cmbx6
\font \sixi            = cmmi6
\font \sixrm           = cmr6
\font \sixsy           = cmsy6
\font \tafonts         = cmbx12
\font \tafontss        = cmbx10
\font \tafontt         = cmbx12  at 17pt
\font \tams            = cmmib10 at 9pt
\font \tenmib          = cmmib10
\font \tamt            = cmmib10 at 17pt
\font \tass            = cmsy10
\font \tasss           = cmsy7
\font \tast            = cmsy10  at 17pt
\font \tbfonts         = cmbx8
\font \tbfontss        = cmbx10  at 6pt
\font \tbfontt         = cmbx12  at 14pt
\font \tbmt            = cmmib10 at 14pt
\font \tbss            = cmsy8
\font \tbsss           = cmsy6
\font \tbst            = cmsy10  at 14pt
\vsize=23.5truecm
\hoffset=-1true cm
\voffset=1true cm
\newdimen\fullhsize
\fullhsize=40cc
\hsize=19.5cc
\def\fullline{\hbox to\fullhsize}
\def\makefootline{\baselineskip=10dd \fullline{\the\footline}}
\def\makeheadline{\vbox to 0pt{\vskip-22.5pt
            \fullline{\vbox to 8.5pt{}\the\headline}\vss}\nointerlineskip}
\let\lr=L \newbox\leftcolumn
\output={\global\topskip=10pt
         \if L\lr
            \global\setbox\leftcolumn=\columnbox \global\let\lr=R
            \message{[left\the\pageno]}%
            \ifnum\pageno=1
               \global\topskip=\fullhead\fi
         \else
            \doubleformat \global\let\lr=L
         \fi
         \ifnum\outputpenalty>-2000 \else\dosupereject\fi}
\def\doubleformat{\shipout\vbox{\makeheadline
    \fullline{\box\leftcolumn\hfil\columnbox}
           \makefootline}
           \advancepageno}
\def\columnbox{\leftline{\pagebody}}
\outer\def\bye{\bigskip\typeset
\sterne=1\ifx\speciali\undefined\else
\loop\smallskip\noindent special character No\number\sterne:
\csname special\romannumeral\sterne\endcsname
\advance\sterne by 1\global\sterne=\sterne
\ifnum\sterne<11\repeat\fi
\if R\lr\null\fi\vfill\supereject\end}
\def\typeset{\begpet\noindent This article was processed by the author using
Sprin\-ger-Ver\-lag \TeX\ AA macro package 1989.\endpet}
\hfuzz=2pt
\vfuzz=2pt
\tolerance=1000
\fontdimen3\tenrm=1.5\fontdimen3\tenrm
\fontdimen7\tenrm=1.5\fontdimen7\tenrm
\abovedisplayskip=3 mm plus6pt minus 4pt
\belowdisplayskip=3 mm plus6pt minus 4pt
\abovedisplayshortskip=0mm plus6pt
\belowdisplayshortskip=2 mm plus4pt minus 4pt
\predisplaypenalty=0
\clubpenalty=20000
\widowpenalty=20000
\parindent=1.5em
\frenchspacing
\def\newline{\hfill\break}%
\nopagenumbers
\def\AALogo{\setbox254=\hbox{ ASTROPHYSICS }%
\vbox{\baselineskip=10dd\hrule\hbox{\vrule\vbox{\kern3pt
\hbox to\wd254{\hfil ASTRONOMY\hfil}
\hbox to\wd254{\hfil AND\hfil}\copy254
\hbox to\wd254{\hfil\number\day.\number\month.\number\year\hfil}
\kern3pt}\vrule}\hrule}}
\def\paglay{\headline={{\tenrm\hsize=.75\fullhsize\ifnum\pageno=1
\vbox{\baselineskip=10dd\hrule\line{\vrule\kern3pt\vbox{\kern3pt
\hbox{\bf A and A Manuskript-Nr.}
\hbox{(will be inserted by hand later)}
\kern3pt\hrule\kern3pt
\hbox{\bf Your thesaurus codes are:}
\hbox{\rightskip=0pt plus3em\advance\hsize by-7pt
\vbox{\noindent\ignorespaces\the\THESAURUS}}
\kern3pt}\hfil\kern3pt\vrule}\hrule}
\rlap{\quad\AALogo}\hfil
\else\ifodd\pageno\hfil\folio\else\folio\hfil\fi\fi}}}
\ifx \undefined\instruct
\headline={\tenrm\ifodd\pageno\hfil\folio
\else\folio\hfil\fi}\fi
\newcount\eqnum\eqnum=0
\def\autnum{\global\advance\eqnum by 1{\rm(\the\eqnum)}}
\newtoks\eq\newtoks\eqn
\catcode`@=11
\def\eqalign#1{\null\vcenter{\openup\jot\m@th
  \ialign{\strut\hfil$\displaystyle{##}$&$\displaystyle{{}##}$\hfil
      \crcr#1\crcr}}}
\def\displaylines#1{{}$\displ@y
\hbox{\vbox{\halign{$\@lign\hfil\displaystyle##\hfil$\crcr
    #1\crcr}}}${}}
\def\eqalignno#1{{}$\displ@y
  \hbox{\vbox{\halign to\hsize{\hfil$\@lign\displaystyle{##}$\tabskip\z@skip
    &$\@lign\displaystyle{{}##}$\hfil\tabskip\centering
    &\llap{$\@lign##$}\tabskip\z@skip\crcr
    #1\crcr}}}${}}
\def\leqalignno#1{{}$\displ@y
\hbox{\vbox{\halign to\hsize{\qquad\hfil$\@lign\displaystyle{##}$\tabskip\z@skip
    &$\@lign\displaystyle{{}##}$\hfil\tabskip\centering
    &\kern-\hsize\rlap{$\@lign##$}\tabskip\hsize\crcr
    #1\crcr}}}${}}
\def\generaldisplay{%
\ifeqno
       \ifleqno\leftline{$\displaystyle\the\eqn\quad\the\eq$}%
       \else\line{$\displaystyle\the\eq\hfill\the\eqn$}\fi
\else
       \leftline{$\displaystyle\the\eq$}%
\fi
\global\eq={}\global\eqn={}}%
\newif\ifeqno\newif\ifleqno \everydisplay{\displaysetup}
\def\displaysetup#1$${\displaytest#1\eqno\eqno\displaytest}
\def\displaytest#1\eqno#2\eqno#3\displaytest{%
\if!#3!\ldisplaytest#1\leqno\leqno\ldisplaytest
\else\eqnotrue\leqnofalse\eqn={#2}\eq={#1}\fi
\generaldisplay$$}
\def\ldisplaytest#1\leqno#2\leqno#3\ldisplaytest{\eq={#1}%
\if!#3!\eqnofalse\else\eqnotrue\leqnotrue\eqn={#2}\fi}
\catcode`@=12 
\mathchardef\Gamma="0100
\mathchardef\Delta="0101
\mathchardef\Theta="0102
\mathchardef\Lambda="0103
\mathchardef\Xi="0104
\mathchardef\Pi="0105
\mathchardef\Sigma="0106
\mathchardef\Upsilon="0107
\mathchardef\Phi="0108
\mathchardef\Psi="0109
\mathchardef\Omega="010A

\def\utw{\smash{\rlap{\lower5pt\hbox{$\sim$}}}}
\def\udtw{\smash{\rlap{\lower6pt\hbox{$\approx$}}}}

\def\diameter{{\ifmmode\mathchoice
{\ooalign{\hfil\hbox{$\displaystyle/$}\hfil\crcr
{\hbox{$\displaystyle\mathchar"20D$}}}}
{\ooalign{\hfil\hbox{$\textstyle/$}\hfil\crcr
{\hbox{$\textstyle\mathchar"20D$}}}}
{\ooalign{\hfil\hbox{$\scriptstyle/$}\hfil\crcr
{\hbox{$\scriptstyle\mathchar"20D$}}}}
{\ooalign{\hfil\hbox{$\scriptscriptstyle/$}\hfil\crcr
{\hbox{$\scriptscriptstyle\mathchar"20D$}}}}
\else{\ooalign{\hfil/\hfil\crcr\mathhexbox20D}}%
\fi}}

\normallineskip=1dd
\normallineskiplimit=0dd
\normalbaselineskip=10dd
\textfont0=\tenrm
\textfont1=\teni
\textfont2=\tensy
\textfont\itfam=\tenit
\textfont\slfam=\tensl
\textfont\ttfam=\tentt
\textfont\bffam=\tenbf
\normalbaselines\rm
\def\petit{\def\rm{\fam0\eightrm}
\textfont0=\eightrm \scriptfont0=\sixrm \scriptscriptfont0=\fiverm
 \textfont1=\eighti \scriptfont1=\sixi \scriptscriptfont1=\fivei
 \textfont2=\eightsy \scriptfont2=\sixsy \scriptscriptfont2=\fivesy
 \def\it{\fam\itfam\eightit}%
 \textfont\itfam=\eightit
 \def\sl{\fam\slfam\eightsl}%
 \textfont\slfam=\eightsl
 \def\bf{\fam\bffam\eightbf}%
 \textfont\bffam=\eightbf \scriptfont\bffam=\sixbf
 \scriptscriptfont\bffam=\fivebf
 \def\tt{\fam\ttfam\eighttt}%
 \textfont\ttfam=\eighttt
 \let\tams=\kleinhalbcurs
 \let\tenbf=\eightbf
 \let\sevenbf=\sixbf
 \normalbaselineskip=9dd
 \if Y\REFEREE \normalbaselineskip=2\normalbaselineskip
 \normallineskip=2\normallineskip\fi
 \setbox\strutbox=\hbox{\vrule height7pt depth2pt width0pt}%
 \normalbaselines\rm}%
\def\begpet{\vskip6pt\bgroup\petit}
\def\endpet{\vskip6pt\egroup}
\def\rahmen#1{\vbox{\hrule\line{\vrule\vbox to#1true
cm{\vfil}\hfil\vrule}\vfil\hrule}}
\def\begfig#1cm#2\endfig{\par
   \ifvoid\topins\midinsert\bigskip\vbox{\rahmen{#1}#2}\endinsert
   \else\topinsert\vbox{\rahmen{#1}#2}\endinsert
\fi}
\def\begfigwid#1cm#2\endfig{\par
\if N\lr\else
\if R\lr
\shipout\vbox{\makeheadline
\line{\box\leftcolumn}\makefootline}\advancepageno
\fi\let\lr=N
\topskip=10pt
\output={\plainoutput}%
\fi
\topinsert\line{\vbox{\hsize=\fullhsize\rahmen{#1}#2}\hss}\endinsert}
\def\figure#1#2{\bigskip\noindent{\petit{\bf Fig.\ts#1.\
}\ignorespaces #2\smallskip}}
\def\begtab#1cm#2\endtab{\par
   \ifvoid\topins\midinsert\medskip\vbox{#2\rahmen{#1}}\endinsert
   \else\topinsert\vbox{#2\rahmen{#1}}\endinsert
\fi}
\def\begtabemptywid#1cm#2\endtab{\par
\if N\lr\else
\if R\lr
\shipout\vbox{\makeheadline
\line{\box\leftcolumn}\makefootline}\advancepageno
\fi\let\lr=N
\topskip=10pt
\output={\plainoutput}%
\fi
\topinsert\line{\vbox{\hsize=\fullhsize#2\rahmen{#1}}\hss}\endinsert}
\def\begtabfullwid#1\endtab{\par
\if N\lr\else
\if R\lr
\shipout\vbox{\makeheadline
\line{\box\leftcolumn}\makefootline}\advancepageno
\fi\let\lr=N
\output={\plainoutput}%
\fi
\topinsert\line{\vbox{\hsize=\fullhsize\noindent#1}\hss}\endinsert}

\def\begref{\vskip1cm\begingroup\let\INS=N}
\def\ref{\goodbreak\if N\INS\let\INS=Y\vbox{\noindent\tenbf
References\bigskip}\fi\hangindent\parindent
\hangafter=1\noindent\ignorespaces}
\def\endref{\goodbreak\endgroup}
\def\ack#1{\vskip11pt\begingroup\noindent{\it Acknowledgements\/}.
\ignorespaces#1\vskip6pt\endgroup}
\def\app#1{\vskip25pt\noindent{\bf Appendix #1}\vskip4.4pt\noindent
\ignorespaces}
%
%
 \def \aTa  { \goodbreak
     \bgroup
     \par
 \textfont0=\tafontt \scriptfont0=\tafonts \scriptscriptfont0=\tafontss
 \textfont1=\tamt \scriptfont1=\tbmt \scriptscriptfont1=\tams
 \textfont2=\tast \scriptfont2=\tass \scriptscriptfont2=\tasss
     \baselineskip=17dd
     \lineskip=17dd
     \rightskip=0pt plus2cm\spaceskip=.3333em \xspaceskip=.5em
     \pretolerance=10000
     \noindent
     \tafontt}
 %
 \def \eTa{\vskip10pt\egroup
     \noindent
     \ignorespaces}
 
%
%
%
 %
 \def \aTb{\goodbreak
     \bgroup
     \par
 \textfont0=\tbfontt \scriptfont0=\tbfonts \scriptscriptfont0=\tbfontss
 \textfont1=\tbmt \scriptfont1=\tenmib \scriptscriptfont1=\tams
 \textfont2=\tbst \scriptfont2=\tbss \scriptscriptfont2=\tbsss
     \baselineskip=13dd
     \lineskip=13dd
     \rightskip=0pt plus2cm\spaceskip=.3333em \xspaceskip=.5em
     \pretolerance=10000
     \noindent
     \tbfontt}
 %
 \def \eTb{\vskip10pt
     \egroup
     \noindent
     \ignorespaces}
  %
\catcode`\@=11
\expandafter \newcount \csname c@Tl\endcsname
    \csname c@Tl\endcsname=0
\expandafter \newcount \csname c@Tm\endcsname
    \csname c@Tm\endcsname=0
\expandafter \newcount \csname c@Tn\endcsname
    \csname c@Tn\endcsname=0
\expandafter \newcount \csname c@To\endcsname
    \csname c@To\endcsname=0
\expandafter \newcount \csname c@Tp\endcsname
    \csname c@Tp\endcsname=0
\def \resetcount#1    {\global
    \csname c@#1\endcsname=0}
\def\@nameuse#1{\csname #1\endcsname}
\def\arabic#1{\@arabic{\@nameuse{c@#1}}}
\def\@arabic#1{\ifnum #1>0 \number #1\fi}
 %
\expandafter \newcount \csname c@fn\endcsname
    \csname c@fn\endcsname=0
\def \stepc#1    {\global
    \expandafter
    \advance
    \csname c@#1\endcsname by 1}
\catcode`\@=12
%
%
   \catcode`\@= 11
%
%
 
\skewchar\eighti='177 \skewchar\sixi='177
\skewchar\eightsy='60 \skewchar\sixsy='60
\hyphenchar\eighttt=-1
\def\footnoterule{\kern-3pt\hrule width 2true cm\kern2.6pt}
\newinsert\footins
\def\footnotea#1{\let\@sf\empty 
  \ifhmode\edef\@sf{\spacefactor\the\spacefactor}\/\fi
  {#1}\@sf\vfootnote{#1}}
\def\vfootnote#1{\insert\footins\bgroup
  \textfont0=\tenrm\scriptfont0=\sevenrm\scriptscriptfont0=\fiverm
  \textfont1=\teni\scriptfont1=\seveni\scriptscriptfont1=\fivei
  \textfont2=\tensy\scriptfont2=\sevensy\scriptscriptfont2=\fivesy
  \interlinepenalty\interfootnotelinepenalty
  \splittopskip\ht\strutbox 
  \splitmaxdepth\dp\strutbox \floatingpenalty\@MM
  \leftskip\z@skip \rightskip\z@skip \spaceskip\z@skip \xspaceskip\z@skip
  \textindent{#1}\footstrut\futurelet\next\fo@t}
\def\fo@t{\ifcat\bgroup\noexpand\next \let\next\f@@t
  \else\let\next\f@t\fi \next}
\def\f@@t{\bgroup\aftergroup\@foot\let\next}
\def\f@t#1{#1\@foot}
\def\@foot{\strut\egroup}
\def\footstrut{\vbox to\splittopskip{}}
\skip\footins=\bigskipamount 
\count\footins=1000 
\dimen\footins=8in 
   \def \bfootax  {\bgroup\tenrm
                  \baselineskip=12pt\lineskiplimit=-6pt
                  \hsize=19.5cc
                  \def\textindent##1{\hang\noindent\hbox
                  to\parindent{##1\hss}\ignorespaces}%
                  \footnotea{$^\star$}\bgroup}
   \def \efootax  {\egroup\egroup}
   \def \bfootay  {\bgroup\tenrm
                  \baselineskip=12pt\lineskiplimit=-6pt
                  \hsize=19.5cc
                  \def\textindent##1{\hang\noindent\hbox
                  to\parindent{##1\hss}\ignorespaces}%
                  \footnotea{$^{\star\star}$}\bgroup}
   \def \efootay  {\egroup\egroup }
   \def \bfootaz {\bgroup\tenrm
                  \baselineskip=12pt\lineskiplimit=-6pt
                  \hsize=19.5cc
                  \def\textindent##1{\hang\noindent\hbox
                  to\parindent{##1\hss}\ignorespaces}%
                 \footnotea{$^{\star\star\star}$}\bgroup}
   \def \efootaz {\egroup \egroup}
\def\fonote#1{\mehrsterne$^{\the\sterne}$\begingroup
       \def\textindent##1{\hang\noindent\hbox
       to\parindent{##1\hss}\ignorespaces}%
\vfootnote{$^{\the\sterne}$}{#1}\endgroup}
\catcode`\@=12
%
\everypar={\let\lasttitle=N\everypar={\parindent=1.5em}}%
%
%
\def \titlea#1{\stepc{Tl}
     \resetcount{Tm}
     \vskip22pt
     \setbox0=\vbox{\vskip 22pt\noindent
     \bf
     \rightskip 0pt plus4em
     \pretolerance=20000
     \arabic{Tl}.\
    \textfont1=\tams\scriptfont1=\kleinhalbcurs\scriptscriptfont1=\kleinhalbcurs
     \ignorespaces#1
     \vskip11pt}
     \dimen0=\ht0\advance\dimen0 by\dp0\advance\dimen0 by 2\baselineskip
     \advance\dimen0 by\pagetotal
     \ifdim\dimen0>\pagegoal\eject\fi
     \bgroup
     \noindent
     \bf
     \rightskip 0pt plus4em
     \pretolerance=20000
     \arabic{Tl}.\
    \textfont1=\tams\scriptfont1=\kleinhalbcurs\scriptscriptfont1=\kleinhalbcurs
     \ignorespaces#1
     \vskip11pt
     \egroup
     \nobreak
     \parindent=0pt
     \everypar={\parindent=1.5em
     \let\lasttitle=N\everypar={\let\lasttitle=N}}%
     \let\lasttitle=A%
     \ignorespaces}
 %
 \def\titleb#1{\stepc{Tm}
     \resetcount{Tn}
     \if N\lasttitle\else\vskip-11pt\vskip-\baselineskip
     \fi
     \vskip 17pt
     \setbox0=\vbox{\vskip 17pt
     \raggedright
     \pretolerance=10000
     \noindent
     \it
     \arabic{Tl}.\arabic{Tm}.\
     \ignorespaces#1
     \vskip8pt}
     \dimen0=\ht0\advance\dimen0 by\dp0\advance\dimen0 by 2\baselineskip
     \advance\dimen0 by\pagetotal
     \ifdim\dimen0>\pagegoal\eject\fi
     \bgroup
     \raggedright
     \pretolerance=10000
     \noindent
     \it
     \arabic{Tl}.\arabic{Tm}.\
     \ignorespaces#1
     \vskip8pt
     \egroup
     \nobreak
     \let\lasttitle=B%
     \parindent=0pt
     \everypar={\parindent=1.5em
     \let\lasttitle=N\everypar={\let\lasttitle=N}}%
     \ignorespaces}
 %
 \def \titlec#1{\stepc{Tn}
     \resetcount{To}
     \if N\lasttitle\else\vskip-3pt\vskip-\baselineskip
     \fi
     \vskip 11pt
     \setbox0=\vbox{\vskip 11pt
     \noindent
     \raggedright
     \pretolerance=10000
     \arabic{Tl}.\arabic{Tm}.\arabic{Tn}.\
     \ignorespaces#1\vskip6pt}
     \dimen0=\ht0\advance\dimen0 by\dp0\advance\dimen0 by 2\baselineskip
     \advance\dimen0 by\pagetotal
     \ifdim\dimen0>\pagegoal\eject\fi
     \bgroup\noindent
     \raggedright
     \pretolerance=10000
     \arabic{Tl}.\arabic{Tm}.\arabic{Tn}.\
     \ignorespaces#1\vskip6pt
     \egroup
     \nobreak
     \let\lasttitle=C%
     \parindent=0pt
     \everypar={\parindent=1.5em
     \let\lasttitle=N\everypar={\let\lasttitle=N}}%
     \ignorespaces}
 %
 \def\titled#1{\stepc{To}
     \resetcount{Tp}
     \if N\lasttitle\else\vskip-3pt\vskip-\baselineskip
     \fi
     \vskip 11pt
     \bgroup
     \it
     \noindent
     \ignorespaces#1\unskip. \egroup\ignorespaces}
\let\REFEREE=N
\newbox\refereebox
\setbox\refereebox=\vbox
to0pt{\vskip0.5cm\fullline{\hrulefill\tentt\lower0.5ex
\hbox{\kern5pt referee's copy\kern5pt}\hrulefill}\vss}%
\def\refereelayout{\let\REFEREE=M\footline={\copy\refereebox}%
\message{|A referee's copy will be produced}\par
\if N\lr\else
\if R\lr
\shipout\vbox{\makeheadline
\line{\box\leftcolumn}\makefootline}\advancepageno
\fi\let\lr=N
\topskip=10pt
\output={\plainoutput}%
\fi
}
\let\ts=\thinspace
\newcount\sterne \sterne=0
\newdimen\fullhead
\newtoks\RECDATE
\newtoks\ACCDATE
\newtoks\MAINTITLE
\newtoks\SUBTITLE
\newtoks\AUTHOR
\newtoks\INSTITUTE
\newtoks\SUMMARY
\newtoks\KEYWORDS
\newtoks\THESAURUS
\newtoks\SENDOFF
\newlinechar=`\| 
\catcode`\@=\active
\let\INS=N%
\def@#1{\if N\INS $^{#1}$\else\if
E\INS\hangindent0.5\parindent\hangafter=1%
\noindent\hbox to0.5\parindent{$^{#1}$\hfil}\let\INS=Y\ignorespaces
\else\par\hangindent0.5\parindent\hangafter=1
\noindent\hbox to0.5\parindent{$^{#1}$\hfil}\ignorespaces\fi\fi}%
\def\mehrsterne{\advance\sterne by1\global\sterne=\sterne}%
\def\FOOTNOTE#1{\mehrsterne\ifcase\sterne
\or\bfootax \ignorespaces #1\efootax
\or\bfootay \ignorespaces #1\efootay
\or\bfootaz \ignorespaces #1\efootaz\else\fi}%
\def\PRESADD#1{\mehrsterne\ifcase\sterne
\or\bfootax Present address: #1\efootax
\or\bfootay Present address: #1\efootay
\or\bfootaz Present address: #1\efootaz\else\fi}%
\def\maketitle{\paglay%
\def\missing{ ????? }
%
\setbox0=\vbox{\parskip=0pt\hsize=\fullhsize\null\vskip2truecm
\let\kka = \tamt
\edef\test{\the\MAINTITLE}%
\ifx\test\missing\MAINTITLE={MAINTITLE should be given}\fi
\aTa\ignorespaces\the\MAINTITLE\eTa
\let\kka = \tbmt
\edef\test{\the\SUBTITLE}%
\ifx\test\missing\else\aTb\ignorespaces\the\SUBTITLE\eTb\fi
\let\kka = \tams
\edef\test{\the\AUTHOR}%
\ifx\test\missing
\AUTHOR={Name(s) and initial(s) of author(s) should be given}\fi
\noindent{\bf\ignorespaces\the\AUTHOR\vskip4pt}
\let\INS=E%
\edef\test{\the\INSTITUTE}%
\ifx\test\missing
\INSTITUTE={Address(es) of author(s) should be given.}\fi
{\noindent\ignorespaces\the\INSTITUTE\vskip10pt}%
\edef\test{\the\RECDATE}%
\ifx\test\missing
\RECDATE={{\petit $[$the date should be inserted later$]$}}\fi
\edef\test{\the\ACCDATE}%
\ifx\test\missing
\ACCDATE={{\petit $[$the date should be inserted later$]$}}\fi
{\noindent Received \ignorespaces\the\RECDATE\unskip; accepted \ignorespaces
\the\ACCDATE\vskip21pt\bf S}}%
\global\fullhead=\ht0\global\advance\fullhead by\dp0
\global\advance\fullhead by10pt\global\sterne=0
{\parskip=0pt\hsize=19.5cc\null\vskip2truecm
\edef\test{\the\SENDOFF}%
\ifx\test\missing\else\insert\footins{\smallskip\noindent
{\it Send offprint requests to\/}: \ignorespaces\the\SENDOFF}\fi
\hsize=\fullhsize
\let\kka = \tamt
\edef\test{\the\MAINTITLE}%
\ifx\test\missing\message{|Your MAINTITLE is missing.}%
\MAINTITLE={MAINTITLE should be given}\fi
\aTa\ignorespaces\the\MAINTITLE\eTa
\let\kka = \tbmt
\edef\test{\the\SUBTITLE}%
\ifx\test\missing\message{|The SUBTITLE is optional.}%
\else\aTb\ignorespaces\the\SUBTITLE\eTb\fi
\let\kka = \tams
\edef\test{\the\AUTHOR}%
\ifx\test\missing\message{|Name(s) and initial(s) of author(s) missing.}%
\AUTHOR={Name(s) and initial(s) of author(s) should be given}\fi
\noindent{\bf\ignorespaces\the\AUTHOR\vskip4pt}
\let\INS=E%
\edef\test{\the\INSTITUTE}%
\ifx\test\missing\message{|Address(es) of author(s) missing.}%
\INSTITUTE={Address(es) of author(s) should be given.}\fi
{\noindent\ignorespaces\the\INSTITUTE\vskip10pt}%
\edef\test{\the\RECDATE}%
\ifx\test\missing\message{|The date of receipt should be inserted
later.}%
\RECDATE={{\petit $[$the date should be inserted later$]$}}\fi
\edef\test{\the\ACCDATE}%
\ifx\test\missing\message{|The date of acceptance should be inserted
later.}%
\ACCDATE={{\petit $[$the date should be inserted later$]$}}\fi
{\noindent Received \ignorespaces\the\RECDATE\unskip; accepted \ignorespaces
\the\ACCDATE\vskip21pt}}%
\edef\test{\the\THESAURUS}%
\ifx\test\missing\THESAURUS={missing; you have not inserted them}%
\message{|Thesaurus codes are not given.}\fi
\if M\REFEREE\let\REFEREE=Y
\normalbaselineskip=2\normalbaselineskip
\normallineskip=2\normallineskip\normalbaselines\fi
\edef\test{\the\SUMMARY}%
\ifx\test\missing\message{|Summary is missing.}%
\SUMMARY={Not yet given.}\fi
\noindent{\bf Summary. }\ignorespaces
\the\SUMMARY\vskip0.5true cm
\edef\test{\the\KEYWORDS}%
\ifx\test\missing\message{|Missing keywords.}%
\KEYWORDS={Not yet given.}\fi
\noindent{\bf Key words: }\the\KEYWORDS
\vskip3pt\line{\hrulefill}\vfill
\global\sterne=0
\catcode`\@=12}


\MAINTITLE={Cosmic Rays}
\SUBTITLE={VII. Individual element spectra:  prediction and data}
\AUTHOR={\ts Barbara Wiebel-Sooth@1, \ts Peter L. Biermann@2, and \ts Hinrich
Meyer@1} \SENDOFF={\ts Peter L. Biermann}
\INSTITUTE={
@1 Fachbereich Physik, Universit\"at Wuppertal, Gau\ss str. 20, D-42097 Wuppertal, Germany,
@2 Max Planck Institut f\"ur Radioastronomie, Auf dem H\"ugel 69,
D-53121 Bonn, Germany}

%
%
%
%
%

\RECDATE={   }
\ACCDATE={   }
\SUMMARY={
Based on the earlier papers in this series, we discuss here the errors in the
prediction of the model, and compare with the error ranges in the spectra for
individual elements.  The predictions are spectra of $E^{-2.75 \pm 0.04}$ for
hydrogen, and $E^{-2.67 -0.02 \pm 0.02}$ for helium and heavier elements below
the knee.  For particle energies above $10 \, Z$ GeV the data give
$E^{-2.77 \pm 0.02}$ for hydrogen  and $E^{-2.64 \pm 0.02}$ for helium and
similar values for all heavier elements combined, where  $Z$ is the charge
of the nucleus considered.  At the higher energy range considered here the
secondary elements Li, Be, and B as well as the sub-Fe  group have spectra
consistent with source-related spallation, such as occurs when a
supernova driven explosions runs into a shell around the wind of the
predecessor star.}

\KEYWORDS={Acceleration of particles -- Cosmic Rays -- Plasmas --
Supernovae: general -- Shockwaves}

\THESAURUS={02.01.1, 09.03.2, 02.16.1, 08.19.4, 02.19.1}

\maketitle

\titlea {Introduction}

The origin of cosmic rays (Hess 1912, Kohlh{\"o}rster 1913) is still one
of the main enigmas of physics.  Many energetic sources have been
claimed to explain parts of the population of cosmic ray particles
observed.  Novae, pulsars, supernova explosions, stellar wind bubbles,
interstellar turbulence, radio galaxies and many other sources have been
explored (Fermi 1949, 1954; Peters 1959, 1961; Berezinsky et al. 1990).

\titleb {The origin of cosmic rays}

Recently we have proposed that cosmic rays can be traced to three different
source sites:

\item 1.  Supernova explosions into the interstellar medium, or ISM-SNe.  This
component produces mostly hydrogen and the observed energetic electrons up to
about 30 GeV, and dominates the {\it all particle} flux up to near $10^4$ GeV.

\item 2.  Supernova explosions into the predecessor stellar wind, wind-SNe.
This component produces the observed energetic electrons above 30 GeV,
and helium and most heavier elements already from GeV particle energies.
Due to a reduction in acceleration efficiency at a particular rigidity
the slope increases, thus producing the knee feature.  The component extends
ultimately to several EeV. Since the winds of massive stars are enriched
late in their life, this component shows a heavy element abundance which
is strongly increased over the interstellar medium.

\item 3.  Powerful radio galaxies produce a contribution which dominates beyond
about 3 EeV, and consists mostly of hydrogen and helium, with only little
addition of heavy elements below 50 EeV.  At this energy the interaction with
the microwave background cuts off the contribution from distant extragalactic
sources, the Greisen-Zatsepin-Kuzmin (GZK) cutoff. 
There are a small number of events which appear to be beyond 
this energy, and whether they fit into such a picture is open at present.

The theory was originally proposed in Biermann (1993a, paper CR I) and in
Rachen \& Biermann (1993, paper UHE CR I).  Various tests were performed in
Biermann \& Cassinelli (1993, paper CR II); Biermann \& Strom (1993, paper CR
III); Stanev et al.  (1993, paper CR IV); Rachen \& Biermann (1993, paper
UHE CR  I); Rachen et al. (1993, paper UHE CR II); Nath \& Biermann (1993,
1994a, 1994b); Biermann et al. (1995a); Biermann et al.
(1995b, paper CR V); Stanev et al. (1995); Biermann et al. (1997) and
Biermann (1993b, c, 1994, 1995a, b, 1996, 1997a, 1997b).

In this paper we will at first briefly discuss the error determination of the
theory in Sect. 2, then the data for individual elements in Sect. 3,
compare and discuss secondary elements versus primary elements in Sect. 4,
and conclude in Sect. 5.

\titlea {Errors inherent in the theory}

\titleb{Basic concept}

We consider the acceleration of particles in spherical shocks with the
concept of first order Fermi acceleration (see, e.g., Drury 1983).  In
this concept energetic particles are cycling back and forth across the shock
region, gaining energy each time they turn back; since the two sides of a
shock are a permanently compressing system the particles gain energy.  In a
spherical shock the particles also lose energy from adiabatic expansion.
Furthermore, in a system where the unperturbed magnetic field is
perpendicular to the shock normal, particles can modify their energy by
drifts, here dominated by curvature drifts; the particles move sideways in a
curved magnetic field, and experience an electric field from the motion
through a magnetic field; the component of the motion parallel to the
perceived electric field leads to an energy change.  It has to be noted that
for plane parallel shocks the drifts have been shown by Jokipii (1982) to be
equivalent in their effect to the Lorentz transformation; here we emphasize
that we use the curvature and gradient drifts only, which give an
additional effect.  A key ingredient in  such an approach is the time  spent
by a particle on either side of the  shock.  Observations as well as stability
arguments lead us to the notion,  that the time spent on either  side of the
shock is given by a transport coefficient $\kappa$ given by fast convective
motion (see Biermann 1993a, 1997b).  The construction of this transport
coefficient is the major step  in the argument, and is based on the concept of
the {\it smallest dominant scale}, a scale either in real space or in phase
space.  The resulting expression for the spectrum of the energetic particles
is the same as in, e.g., Drury (1983), except for an additional term for energy
gains from drifts (Jokipii 1987).

For the expansion of a spherical shock into a stellar wind we adopt the basic
magnetic configuration of a Parker spiral (Parker 1958, Jokipii et al.
1977), where the magnetic field in the equatorial plane is an Archimedian
spiral with $B_{\phi}$ dominating over  $B_r$, decreases with $B_{\phi} \sim
(\sin \theta)/r$ towards the pole and outwards with radius $r$,  and becomes
mostly  radial in the pole region.   We write for the wind velocity $V_W$, for
the shock velocity $U_1$, and for  the downstream gas velocity in the shock
frame $U_2$.

\titleb {Assumptions and Systematic Uncertainties}

The assumptions adopted are inspired by Prandtl{'}s mixing length approach
(Prandtl 1925, 1949); all use the key proposition that the
{\bf smallest dominant scale}, either in geometric length, or in velocity
space, gives the natural transport coefficient.  In this sense the
assumptions are {\it derived from a basic principle}.

Our basic, {\bf argument 1}, {\it based on observational
evidence as well as theoretical arguments}, is that for a cosmic ray
mediated shock the {\it convective random walk} of energetic particles
perpendicular to the {\it unperturbed} magnetic  field can be described
by a diffusive process with a downstream diffusion coefficient
$\kappa_{rr,2}$ which is given by the thickness of the shocked layer and the
velocity difference across the shock, and is independent of energy.

The upstream diffusion coefficient can be derived in a similar way, but with
a larger scale based on the {\it same column density as in the downstream
layer}.  This leads to the second critical, {\bf argument 2}, namely that the
upstream length scale is just $U_1/U_2$ times larger.

It must be remembered that there is a lot of convective turbulence which
increases the curvature:  The characteristic scale of the turbulence is
$r/4$ for strong shocks, again, as an example, in the case of the wind-SN,
and thus the curvature is $4/r$ maximum.  Half the maximum of the curvature
allows for the net balance of gains and losses for the energy gain due
to drifts ({\bf argument 3}), and so we obtain then for the curvature
$2/r$ which is twice the curvature without any turbulence; this increases
the curvature term for the spectral range below the knee.

The diffusion tensor component $\kappa_{\theta \theta}$
can be derived similar to our heuristic derivation of the radial diffusion
term $\kappa_{r r}$, again by using the smallest dominant scales.  The
characteristic velocity of particles in $\theta$ is given by the erratic
part of the drifting, corresponding to spatial elements of different
magnetic field direction, and this is on average the value of the drift
velocity  $\mid V_{d,\theta}\mid$, possibly modified by the locally
increased values of the magnetic field strength, and the characteristic
length is the distance to the symmetry axis $r\,sin\,\theta$ ({\bf
argument 4});  this is the smallest dominant scale as soon as the thickness
of the shocked  layer is larger than the distance to the symmetry axis, 
i.e.  $sin \,\theta\,<U_2 /U_1$.

Rapid convection also gives a competing diffusion in the $\theta$-direction,
independent of particle energy; this will begin to dominate as soon as the
energy  dependent $\theta$-diffusion coefficient reaches this maximum at a
critical energy.  As long as the $\theta$-diffusion coefficient is smaller,
it will dominate particle transport in $\theta$ and the upper limit derived
here is irrelevant. When the $\theta$-diffusion coefficient reaches and
passes this maximum given by the fast convection,  then the particle in its
drift will no longer see an increased curvature due to the
convective turbulence due to averaging and the part of drift acceleration
due to increased curvature is eliminated.  Again, a detailed consideration
of gains and losses of the drift energy gains leads to the spectrum of
particles beyond the knee.  The critical energy derived in this way is the
same as that derived from a phase-space argument near the poles.

All these arguments are inspired by Prandtl's mixing length approach; all
use the key proposition that the {\bf smallest dominant scale}, either in
geometric length, or in velocity space, gives the diffusive transport
discussed.  We assume this to be true even for the anisotropic transport
parallel and perpendicular to the shock.

We emphasize the analogy to simple limiting scaling arguments such as
a) the estimate of the temperature gradient in the lower hydrogen convection
zone on the Sun, which is followed by nature to a very good approximation
(Str{\"o}mgren 1953, p. 65), and b) the estimate of the Kolmogorov turbulence
spectrum (Rickett 1990, Goldstein et al. 1995), which appears to be also
followed by nature in many sites over many orders of magnitude in length 
scale. Whether the cosmic rays follow also such a limiting scaling argument, 
as regards their spectrum, to such an accuracy remains to be seen. This 
paper is a step to verify the straight forward prediction for 28 chemical
elements individually.

In addition, we i) use the simplified notion of a purely spherical shock;
ii) ignore the modifications of the shock introduced by the cosmic rays
themselves, except in the conceptual derivation of the initial argument,
where the cosmic rays are critical for the instability; and iii) use a test
particle approach.

We have to emphasize very strongly that these uncertainties mean that the
spectral indices derived for the powerlaw region of the various components
of the cosmic rays correspond to a limiting argument:  If things were
really as simple - and they are likely to be much more complicated - then the
spectrum derived and any comparison with data has to be taken with
considerable caution.

On the other hand, the very simplicity of the proposed concept makes it
easier to test and this is what we propose to do.

\titleb{The error budget}

\titlec {Below the knee}

As discussed in paper CR III, the simplifications which we did make in treating
the flow field of the expansion of a supernova explosion into the interstellar
medium lead to an uncertainty of $\pm 0.04$ in spectral index.

For wind-supernovae we can estimate one uncertainty, which arises from the
finite wind speed of Wolf Rayet stars, or those massive stars with strong winds
which explode as supernovae.  These wind speeds can go up to several
thousand km/sec, while  the supernova shock is variously estimated to $10^4$
km/sec to twice that  much.  As a limiting argument we use that the ratio of
the wind speed to the supernova shock speed is $< 0.2 $; this gives a
steepening of the derived spectral index of the particle distribution by
$0.04$.  This uncertainty also  may correspond to curvature of the spectrum,
since there is a time-evolution  as the shock progresses out through the
stellar wind:  As more energy of the shock is dissipated and more mass of the
stellar wind snowplowed, the shock slows down; then those particles already
accelerated keep their flatter  spectrum (see Eq. 2.44 of Drury 1983), while
those particles freshly injected  and accelerated will have a steeper
spectrum.  Thus, in the range $V_W/U_1 = 0.,..., 0.2$ we obtain a spectral
index in the range $7/3,..., 7/3+0.04$.  Therefore we ascribe to the spectral
index derived here an uncertainty of  $0.02 \pm 0.02$, which describes both
the uncertainty in an assumed powerlaw, and the possible curvature.  This
way of writing makes it clear that we {\it  do not} expect the distribution
of observed spectral indices to follow a  gaussian distribution, but rather
to be biased towards faster winds, which steepen the spectrum.

\titlec {The knee}

In the Fermi-acceleration process there is energy gain and energy loss in each
cycle which an energetic particle crosses the shock region; one part of this
energy gain is due to drifts.  At a certain rigidity ($\sim E/Z$) the drift
contribution is reduced, and so the slope of the spectrum changes.  This
critical energy is given by

$$E_{knee}\; =\;Z e B(r) r ({3 \over 4} {U_1 \over c})^2. \eqno\autnum $$

\noindent In a Parker-wind the product of radius $r$ and magnetic field
strength $B(r)$ is a constant with radius.  $Z$ is the charge of the particle
and $e$ is the electromagnetic charge unit.  $U_1$ is the advance speed of the
shock caused by the supernova explosion; $U_2 = U_1/4$.

This implies that the chemical composition at the knee changes
so, that the gyroradius of the particles at the spectral break is the same,
implying that the different nuclei break off in order of their charge $Z$,
considered as particles of a certain energy (and not as energy per
nucleon).  In an all-particle spectrum in energy per particle, this 
introduces a considerable smearing.

There is one additional cosmic ray component from that latitude region near
the pole of the magnetic field structure in the wind, where the magnetic
field is predominantly radial rather than tangential.  This region we call
the polar cap.  Thus the spectrum is harder in the polar cap region,
because we are close to the standard parallel shock configuration, for which
the particle spectrum is well approximated by $E^{-2}$ at the source.
Because of spatial limitations most of the hemisphere has to dominate again
above the knee, although with a fraction of the  hemisphere that decreases
with particle energy. This introduces a weak progressive steepening of the
spectrum with energy. The superposition of such spectra for different
chemical  elements, including the polar cap contribution, has been tested
(see paper CR IV).  The results of these checks suggest that the polar cap
may be the source for the flattening of the cosmic ray spectrum as one
approaches the knee feature.

We note that we are using a limiting argument to derive the spectrum below
the knee, and again use a limiting argument (see below) for the spectrum
above the knee.  Close to the knee, such an argument breaks down on either
side, and so a softening of the knee feature is to be expected.  On top of
such a softened knee feature the polar cap is an additional component.

The expression for the particle energy at the knee also suggests by the
clearly observed break of the spectrum that the actual values
of the combination $B(r) r U_1^2$ must be limited in range for all supernovae
that contribute appreciably in this energy range.  We have speculated on
possible reasons for such a behaviour elsewhere (paper CR I, and in Biermann
1995a).  We note that the recent results from the Tibet array suggest that 
the knee may in fact be rather smooth (Amenomori et al. 1996), in 
apparent contrast to the earlier description of the Akeno data (Nagano et
al. 1984; see also Stanev et al. 1993).

\titlec {Beyond the knee}

Beyond the knee, the drift contribution to the cyclical energy gain of
individual particles is reduced, and so we obtain a steeper spectrum with an
error which we write as $0.07 \pm 0.07$.  As noted above the use of limiting
arguments to derive the spectrum on either side of the knee implies that the
knee itself may be quite soft, and thus curvature is to be expected.

\titlec {The ultimate cutoff}

The maximum energy particles can reach depends linearly on the magnetic field

$$E_{max} \; = \; Z e r B(r)  .\eqno\autnum $$

If stars that explode as wind-supernovae were to vary widely in their magnetic
field strength, then this maximum energy would also vary from star to star, and
as a result the sum of all contribution would appear as strongly curved
downwards.

\titleb {The predictions}

The proposal is that three sites of origin account for the cosmic rays
observed, i) supernova explosions into the interstellar medium, ISM-SN, ii)
supernova explosions into the stellar wind of the predecessor star, wind-SN,
and iii) radio galaxy hot spots for the extragalactic component.  Here the
cosmic rays attributed to supernova-shocks in stellar winds, wind-SN, produce an
important contribution at all energies up to $3 \, 10^9$ GeV.

Particle energies go up to $100$ Z TeV for ISM-SN, and to $100$ Z PeV with a
bend at $600$ Z TeV for wind-SN.  Radiogalaxy hot spots may go up to near
$1000$ EeV at the source.  These numerical values are estimates with
uncertainties of surely larger than a factor of $2$, since they derive from
an estimated strength of the magnetic field, and estimated values of the
effective shock velocity (see above).

The spectra $\Phi (E)$ are predicted to be

$$\Phi_{ISM} (E) \; \sim \; E^{-2.75 \pm 0.04} \eqno\autnum $$

\noindent for ISM-SN (paper CR III), and

$$\Phi_{wind} (E) \; \sim \; E^{-2.67 - 0.02 \pm 0.02} \eqno\autnum $$

\noindent for wind-SN below the knee, $E^{-3.07 - 0.07 \pm 0.07}$ for
wind-SN above the knee,  and $E^{-2.0}$ at injection for radiogalaxy hot spots.
The polar cap of the wind-SN contributes an $E^{-2.33}$ component (allowing for
leakage from the Galaxy), which, however, contributes significantly only near
and below the knee, if at all.

The chemical abundances are near normal for the injection from ISM-SN, and are
strongly enriched for the contributions from wind-SN. 

This means that the sources for cosmic ray particles and the sources for the
enrichment of the interstellar medium are the same, and hence it is no surprise
that the isotopic ratios are similar for galactic cosmic ray sources and the
solar system (DuVernois et al. 1996a, 1996b).

That spallation-produced
secondaries require a strongly enriched original composition has been confirmed
by a calculation of the formation of light elements in the early Galaxy and a
comparison with observed abundances (Ramaty et al. 1997); we consider this
to be an important check on the picture proposed here.  

At the knee the
spectrum bends downwards at a given rigidity, and so the heavier elements bend
downwards at higher energy per particle.  Thus beyond the knee the heavy
elements dominate all the way to the switchover to the extragalactic component,
which is, once again, mostly Hydrogen and Helium, corresponding to what is
expected to contribute from the interstellar medium of a radiogalaxy, as well as
from any intergalactic contribution mixed in (Biermann 1993c).  This continuous
mix in the chemical composition at the knee already renders the overall knee
feature in a spectrum in energy per particle unavoidably quite smooth, a
tendency which can only partially be offset by the possible polar cap
contribution, since that component also is strongest at a given rigidity (for
details see the discussion in paper CR IV). This is confirmed by Amenomori
et al. (1996). They determined a spectral index of $-2.60 \pm 0.04$ below the
knee and $-3.00 \pm 0.05$ above the knee with the slope changing continuously
between $10^{14.75}$ eV and $10^{15.85}$  eV.

We note that uncertainties of the spectrum derive from a) the time evolution
of any acceleration process as the shock races outward, b) the match between
ISM-SN and wind-SN, c) the mixing of different stellar sources with possibly
different magnetic properties, and d) the differences in propagation in any
model which uses different source populations.  These uncertainties translate
into a distribution of powerlaw indices of the spectra, to curvature of the
spectra, to a smearing of the knee feature, and to a smoothing of the
cutoffs.  Obviously, this is in addition to the underlying uncertainty
associated with the concept of the {\it smallest dominant scale} itself.

\titlea {The data}

Here we describe the basic data set as available in the literature
for individual elements, give several plots, and tables.
The sources for the data are given in Table 4 of the Appendix.
The spectra have been fitted above an energy of $10 \, Z$
GeV, where $Z$ is the charge of the nucleus, in order to minimize the effect
of solar modulation; we repeated the exercise with $20$ and $30 \, Z$ GeV as 
the lower
boundary without finding a significant change (see, e.g.
Garcia-Munoz et al. 1986; Evenson \& Meyer 1984; Seo et al. 1991).  
The numbers given in the tables are from the
first boundary above.  For the fit we have used the CERN-library (MINUIT -
Function Minimization and Error Analysis, CERN Program Library D506).  The
spectra fitted are of the form

$$\Phi \; = \; \Phi_o \, (E/{\rm TeV})^{-\gamma}
{\rm (m^2 \, sec \, sr \, TeV)^{-1}},\eqno\autnum$$
where $E$ is the energy per particle.

$$
$$

\vbox{
{\centerline{\bf Table 1}} \vskip3pt
\vbox{\vfil\hrule
                     \hbox to \hsize{\vrule\hfill
                                     \vbox{\kern10pt
{\settabs \+ Elementy & 222 & $(10.91 \pm 0.32) \;
10^{-222} $ & $ x2.75 \pm 0.222$& 222222 & \cr \+ Element    &\hfill Z
 &\hfil $\Phi_o$ &\hfil $\gamma$ &\hfill $\chi^2/df$ &\cr

\+ H    &\hfill 1   &\hfill $(11.51 \pm 0.41)\; 10^{-2}$  &\hfill
$2.77 \pm 0.02$ &\hfill 0.70 &\cr
\+ He   &\hfill 2   &\hfill $(7.19 \pm 0.20)\;  10^{-2}$  &\hfill
$2.64 \pm 0.02$ &\hfill 2.63 &\cr
\+ Li   &\hfill 3   &\hfill $(2.08 \pm 0.51)\;  10^{-3}$  &\hfill
$2.54 \pm 0.09$ &\hfill 0.90 &\cr
\+ Be   &\hfill 4   &\hfill $(4.74 \pm 0.48)\;  10^{-4}$ &\hfill
$2.75 \pm 0.04$ &\hfill 0.37 &\cr
\+ B    &\hfill 5   &\hfill $(8.95 \pm 0.79)\;  10^{-4}$ &\hfill
$2.95 \pm 0.05$ &\hfill 0.45 &\cr
\+ C    &\hfill 6   &\hfill $(1.06 \pm 0.01)\;  10^{-2}$ &\hfill
$2.66 \pm 0.02$ &\hfill 1.42 &\cr
\+ N    &\hfill 7   &\hfill $(2.35 \pm 0.08)\;  10^{-3}$ &\hfill
$2.72 \pm 0.05$ &\hfill 1.91 &\cr
\+ O     &\hfill 8  &\hfill $(1.57 \pm 0.04)\;  10^{-2}$ &\hfill
$2.68 \pm 0.03$ &\hfill 1.70 &\cr
\+ F    &\hfill 9   &\hfill $(3.28 \pm 0.48)\;  10^{-4}$ &\hfill
$2.69 \pm 0.08$ &\hfill 0.47 &\cr
\+ Ne   &\hfill 10  &\hfill $(4.60 \pm 0.10)\;  10^{-3}$ &\hfill
$2.64 \pm 0.03$ &\hfill 3.14 &\cr
\+ Na   &\hfill 11  &\hfill $(7.54 \pm 0.33)\;  10^{-4}$ &\hfill
$2.66 \pm 0.04$ &\hfill 0.36 &\cr
\+ Mg   &\hfill 12  &\hfill $(8.01 \pm 0.26)\;  10^{-3}$ &\hfill
$2.64 \pm 0.04$ &\hfill 0.10 &\cr
\+ Al   &\hfill 13  &\hfill $(1.15 \pm 0.15)\;  10^{-3}$ &\hfill
$2.66 \pm 0.04$ &\hfill 1.24 &\cr
\+ Si   &\hfill 14  &\hfill $(7.96 \pm 0.15)\;  10^{-3}$ &\hfill
$2.75 \pm 0.04$ &\hfill 0.10 &\cr
\+ P    &\hfill 15  &\hfill $(2.70 \pm 0.20)\;  10^{-4}$ &\hfill
$2.69 \pm 0.06$ &\hfill 0.68 &\cr
\+ S    &\hfill 16  &\hfill $(2.29 \pm 0.24)\;  10^{-3}$ &\hfill
$2.55 \pm 0.09$ &\hfill 0.44 &\cr
\+ Cl   &\hfill 17  &\hfill $(2.94 \pm 0.19)\;  10^{-4}$ &\hfill
$2.68 \pm 0.05$ &\hfill 2.36 &\cr
\+ Ar   &\hfill 18  &\hfill $(8.36 \pm 0.38)\;  10^{-4}$ &\hfill
$2.64 \pm 0.06$ &\hfill 0.45 &\cr
\+ K    &\hfill 19  &\hfill $(5.36 \pm 0.15)\;  10^{-4}$ &\hfill
$2.65 \pm 0.04$ &\hfill 4.58 &\cr
\+ Ca   &\hfill 20  &\hfill $(1.47 \pm 0.12)\;  10^{-3}$ &\hfill
$2.70 \pm 0.06$ &\hfill 0.60 &\cr
\+ Sc   &\hfill 21  &\hfill $(3.04 \pm 0.19)\;  10^{-4}$ &\hfill
$2.64 \pm 0.06$ &\hfill 0.81 &\cr
\+ Ti   &\hfill 22  &\hfill $(1.13 \pm 0.14)\;  10^{-3}$ &\hfill
$2.61 \pm 0.06$ &\hfill 5.67 &\cr
\+ V    &\hfill 23  &\hfill $(6.31 \pm 0.28)\;  10^{-4}$ &\hfill
$2.63 \pm 0.05$ &\hfill 6.83 &\cr
\+ Cr   &\hfill 24  &\hfill $(1.36 \pm 0.12)\;  10^{-3}$ &\hfill
$2.67 \pm 0.06$ &\hfill 3.41 &\cr
\+ Mn   &\hfill 25  &\hfill $(1.35 \pm 0.14)\;  10^{-3}$ &\hfill
$2.46 \pm 0.22$ &\hfill 5.38 &\cr
\+ Fe   &\hfill 26  &\hfill $(1.78 \pm 0.18)\;  10^{-2}$ &\hfill
$2.60 \pm 0.09$ &\hfill 1.81 &\cr
\+ Co   &\hfill 27  &\hfill $(7.51 \pm 0.37)\;  10^{-5}$ &\hfill
$2.72 \pm 0.09$ &\hfill 1.13 &\cr
\+ Ni   &\hfill 28  &\hfill $(9.96 \pm 0.43)\;  10^{-4}$ &\hfill
$2.51 \pm 0.18$ &\hfill 5.47 &\cr} \kern10pt}
                                     \hfill\vrule}
                     \hrule\vfil}
}

$$
$$
We note that the spectral fit takes the experimental errors as given by the
original authors into full account. The fairly small final error is just due
to the assembly of all existing data and their critical combination.
We have divided the results into three tables.  Table 1 gives the results from
experiments where the elements can be separated.  Table 2 gives the results
where groups of elements have been measured together.  Table 3 gives the
results where we have added the data for groups of elements together, both
based on individual measurements  as well as from grouped measurements.
In addition the spectral indices for the various elements as a function of
nuclear charge number $Z$ are shown in Fig. 1.

The data for He illustrate systematic errors in the normalization of the flux:
the lower energy data by Seo et al. (1991) give a spectral index of $2.68 \pm 0.03$,
here we obtain $2.64 \pm 0.02$, which is quite consistent.  The graphs in
Biermann et al. (1995a) clearly show that the slopes of the spectra of He for
individual measurement campaigns agree, while their overall normalization is
different.  When combining data  from different measurement techniques and
campaigns this can lead to erroneous spectral indices, since different
experiments cover different energy ranges. For example, when we apply the fit
by omitting the data of Ryan et al. (1972), we obtain a spectral index for He
of $2.65 \pm 0.02$.

$$
$$

\vbox{
{{\centerline{\bf Table 2}} \vskip3pt
\vbox{\vfil\hrule
                     \hbox to \hsize{\vrule\hfill
                                     \vbox{\kern10pt
{\settabs \+ Elementy & 222 & $x(10.91 \pm 0.32) \;
10^{-222} $ & $x2.75 \pm 0.222$&  222222 & \cr
\+ Element    &\hfill Z        &\hfil $\Phi_o$         &\hfil $\gamma$
&\hfill $\chi^2/df$&\cr

\+ Ne - S  &\hfill 10 - 16  &\hfill $(3.34 \pm 0.18)\;10^{-2}$ &\hfill
$ 2.69  \pm 0.03$ &\hfill 4.02 &\cr
\+ Cl - Ca  &\hfill 17 - 20  &\hfill $(3.35 \pm 0.15)\;10^{-3}$ &\hfill
$ 2.74  \pm 0.03$ &\hfill 0.35 &\cr
\+ Sc - Mn  &\hfill 21 - 25  &\hfill $(4.23 \pm 0.10)\;10^{-3}$ &\hfill
$ 2.77 \pm 0.04 $ &\hfill 2.04 &\cr
\+ Fe - Ni  &\hfill 26 - 28  &\hfill $(1.97 \pm 0.10)\;10^{-2}$ &\hfill
$ 2.62 \pm 0.03$ &\hfill 0.12 &\cr
}\kern10pt}
                                     \hfill\vrule}
                     \hrule\vfil}}
}
$$
$$

Table 2 shows the results of the fits in the case when experiments measured 
groups of
elements directly.  The comparison of Table 1 and 2 illustrates systematic
errors still inherent in the data; for instance the individual elements Cl
to Ca have all been measured by Engelmann et al. (1985, 1990), 
with Ca also measured by Ichimura et al. (1993a, b, c) and Kawamura
et al. (1990). 
The group of elements Cl to Ca has been measured by Simon et al. (1980),
Asakimori et al. (1991), Burnett et al. (1990a, b), JACEE (1993), and 
Ichimura et al. (1993a, b, c).  
Thus the two data sets are independent and allow
to estimate errors.  The difference illustrates that the systematic errors
can be larger than the statistical errors, and that any agreement between
spectral indices of prediction and data, as well as data for different
elements has to be taken with some caution.  In view of the diversity of the
various experiments that contribute to the individual element spectra and the
difficulties to give correct estimates of the systematic uncertainties, we
consider the values for $\chi^2/df$ obtained for the simple powerlaw fits as
quite satisfactory.  With this in view we combine elements to determine the
power law parameters for groups of elements as shown in Table 3.

Table 3 shows the fits to all data available for the various groupings.
We note that the particle spectra are close to the prediction:  for hydrogen
the prediction was $2.75 \pm 0.04$, and the data give $2.77 \pm 0.02$; for
the heavier elements He through Ni the prediction was $2.67 +0.02 \pm 0.02$,
and the data give $2.64 \pm 0.04$.


\vbox{
{{\centerline{\bf Table 3}}\vskip3pt
\vbox{\vfil\hrule
                     \hbox to \hsize{\vrule\hfill
                                     \vbox{\kern10pt
{\settabs \+ Elementyyy & 222222 & $x(10.91 \pm 0.32) \;
10^{-222} $ & $x2.75 \pm 0.222$& \cr
\+ Element    &\hfill Z       &\hfil $\Phi_o$   &\hfil $\gamma$ &\cr

\+ low        &\hfill 1 - 2    &\hfill $(18.81 \pm 0.76)\;10^{-2} $ &\hfill
$2.71 \pm 0.02 $ &\cr
\+ Li - B     &\hfill 3 - 5    &\hfill $(3.49 \pm 0.40)\;10^{-3}  $ &\hfill
$ 2.72 \pm 0.06$ &\cr
\+ Be + B     &\hfill 4 - 5    &\hfill $(1.36 \pm 0.11)\;10^{-3}  $ &\hfill
$ 2.90 \pm 0.04$ &\cr
\+ medium     &\hfill 6 - 8    &\hfill $(2.86 \pm 0.06)\;10^{-2}  $ &\hfill
$2.67 \pm 0.02 $ &\cr
\+ high       &\hfill 10 - 16  &\hfill $(2.84 \pm 0.19)\;10^{-2}  $ &\hfill
$2.66 \pm 0.03 $ &\cr
\+ very high  &\hfill 17 - 26  &\hfill $(1.34 \pm 0.09)\;10^{-2}  $ &\hfill
$2.63 \pm 0.03 $ &\cr
\+ Sc - Mn    &\hfill 21 - 25  &\hfill $(4.74 \pm 0.20)\;10^{-3}  $ &\hfill
$ 2.63 \pm 0.03 $ &\cr
\+ Fe - Ni    &\hfill 26 - 28  &\hfill $(1.89 \pm 0.10)\;10^{-2}  $ &\hfill
$ 2.60 \pm 0.04$ &\cr
\+ He - Ni    &\hfill 2 - 28  &\hfill $(15.80 \pm 1.35)\;10^{-2}  $ &\hfill
$ 2.64 \pm 0.04$ &\cr
\+ allparticle &\hfill 1 - 28  &\hfill $(27.47 \pm 2.43)\;10^{-2} $ &\hfill
$ 2.68 \pm 0.02$ &\cr}\kern10pt}
                                     \hfill\vrule}
                     \hrule\vfil}}
}

$$
$$

We conclude that despite the many simplifications in the theoretical approach
the data are quite consistent with the predictions.

\begfig 11cm
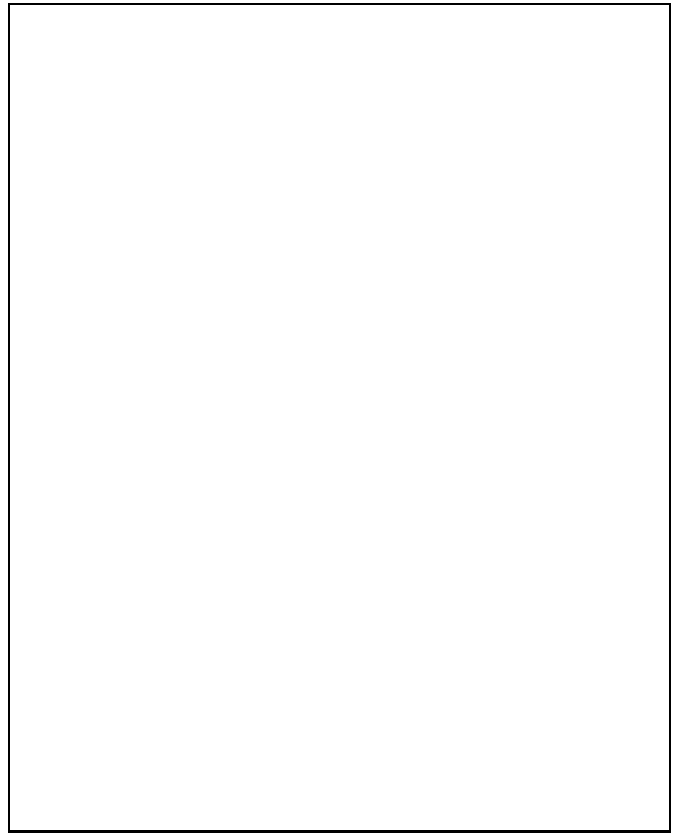
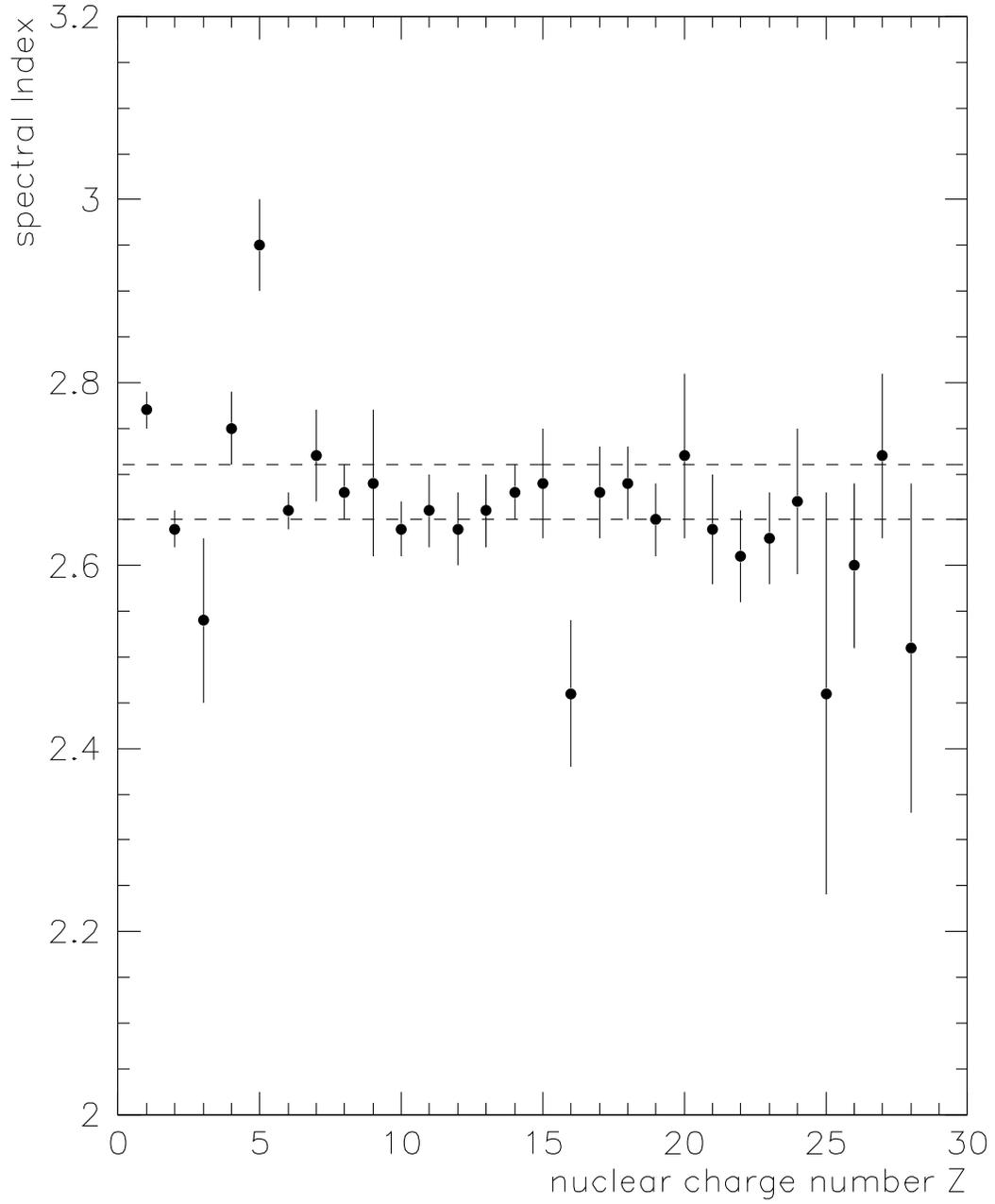
\figure{1}{The spectral indices for the various elements as a function of the
nuclear charge number $Z$. The dashed lines indicate the range of the spectral
index predicted for wind-SN.}
\endfig

We note that for the Fe-group elements we have {\it ignored}
here the spallation correction from interaction in the interstellar medium,
which produces a slight flattening; although not significant, the data are
consistent with such a moderate flattening of these nuclei.

\titlea{The spectra of secondaries}

Secondaries are produced when heavy nuclei break up in collisions with thermal
nuclei, mostly hydrogen in the interstellar medium.  If the production is
throughout the time of transport then the secondaries are expected to show a
steeper spectrum; if on the other hand the spallation is mostly already in
the source region, then the secondary spectrum ought to be the same as that
of the primaries.

The theory for the transport and the production of secondaries is well
developed (e.g. Garcia-Munoz et al. 1987), and uses the steady state
leaky box model to  describe the spectrum as well as the secondary production;
a recent review and summary of the problems inherent in this endeavor has
been given by Shibata (1995).  In a steady state leaky box model the ratio
of the spectra of the secondary elements to the primary elements such as the
boron/carbon or the sub-Fe/Fe ratio directly give the energy dependence of
the diffusion time scale out of the Galaxy.  Such a calculation gives an
energy dependence for the time scale of $\sim \, E^{-0.6}$ (e.g. Engelmann
et al. 1990); one difficulty with this result is that it would lead
to anisotropies of the cosmic rays and would not allow a direct
extrapolation to energies beyond the knee.  There is no other evidence for
the relatively small scales in the interstellar medium of any turbulence
other than consistent with a Kolmogorov spectrum, and also there is no
evidence for a particular length scale corresponding to the Larmor radius
at the knee.  Therefore, we have proposed (Biermann 1996, 1997b), that the
production of secondaries has three independent stages:

a)  First of all, the giant molecular clouds themselves have a time evolution
which cannot be neglected; they form out of smaller cloudlets faster than the
Alfv{\'e}nic time scale, and so {\it trap} the cosmic ray particles.  These
particles then leak out, with an energy dependence given by the Alfv{\'e}nic
turbulence in the cloud.  We note that this corresponds to length scales far
below what can be directly inferred from high angular resolution observations
in clouds.  Assuming that this turbulence also corresponds to a
Kolmogorov spectrum (see, e.g., Goldstein et al. 1995), we then have a
production rate of secondaries, which itself is a function of energy, and
decreases with energy as $E^{-1/3}$.  These particles are then released
into the
interstellar medium, where they are subject to diffusion out of the
Galaxy, and
so acquire another energy dependence of again $E^{-1/3}$, so as to
give a total
energy dependence of the secondary to primary ratio of $E^{-2/3}$,
very close
to that observed.  This is of course limited to that energy range where
diffusion is a useful concept in clouds, and when the time scale of
diffusion
is actually significantly shorter than the lifetime of the giant molecular
cloud.  Thus there is a critical energy per particle, which we
estimate to near
20 $Z$ GeV, above which this process becomes irrelevant.  The detailed
analytical derivation of this argument is given in full in
Biermann (1996), and
will be expanded upon in further communications.

b)  Above this critical energy of an estimated 20 $Z$ GeV, the Galaxy
and its
molecular clouds behave as stationary targets for cosmic ray
interaction, and
we come back to the canonical model, such as explained in
Garcia-Munoz et al. (1987).  
Therefore, for these particles the secondary to primary ratio
just acquires the energy dependence of interstellar turbulence, and so the
ratio is expected to be $\sim \, E^{-1/3}$.  This dependence is a smooth
continuation of the steeper dependence at lower energies.

c)  There is an additional contribution, which arises from the
source, and
which should be energy independent:  As shown in Nath \&
Biermann (1994b)
supernovae that explode into winds, hit a surrounding molecular shell,
and then produce secondaries with an approximate grammage of order 1 - 3
gm/cm$^2$ (= column density traversed in the zig-zag path of
charged particles
in an inhomogeneous magnetic field).  This then leads to a ratio
of secondaries
to primary particles which is energy independent, an aspect that has also
been remarked by others (Drury et al. 1993).

\begfig 11cm
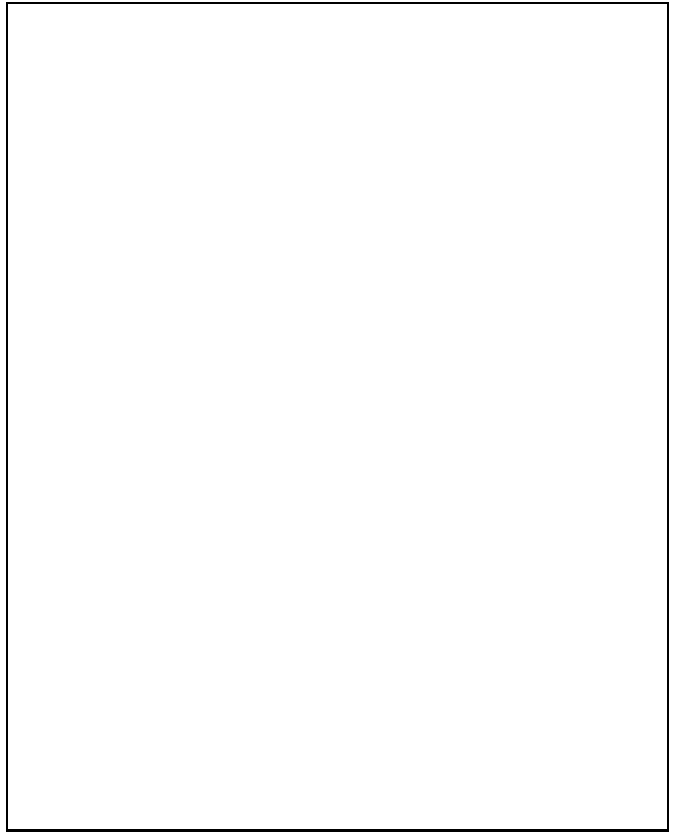
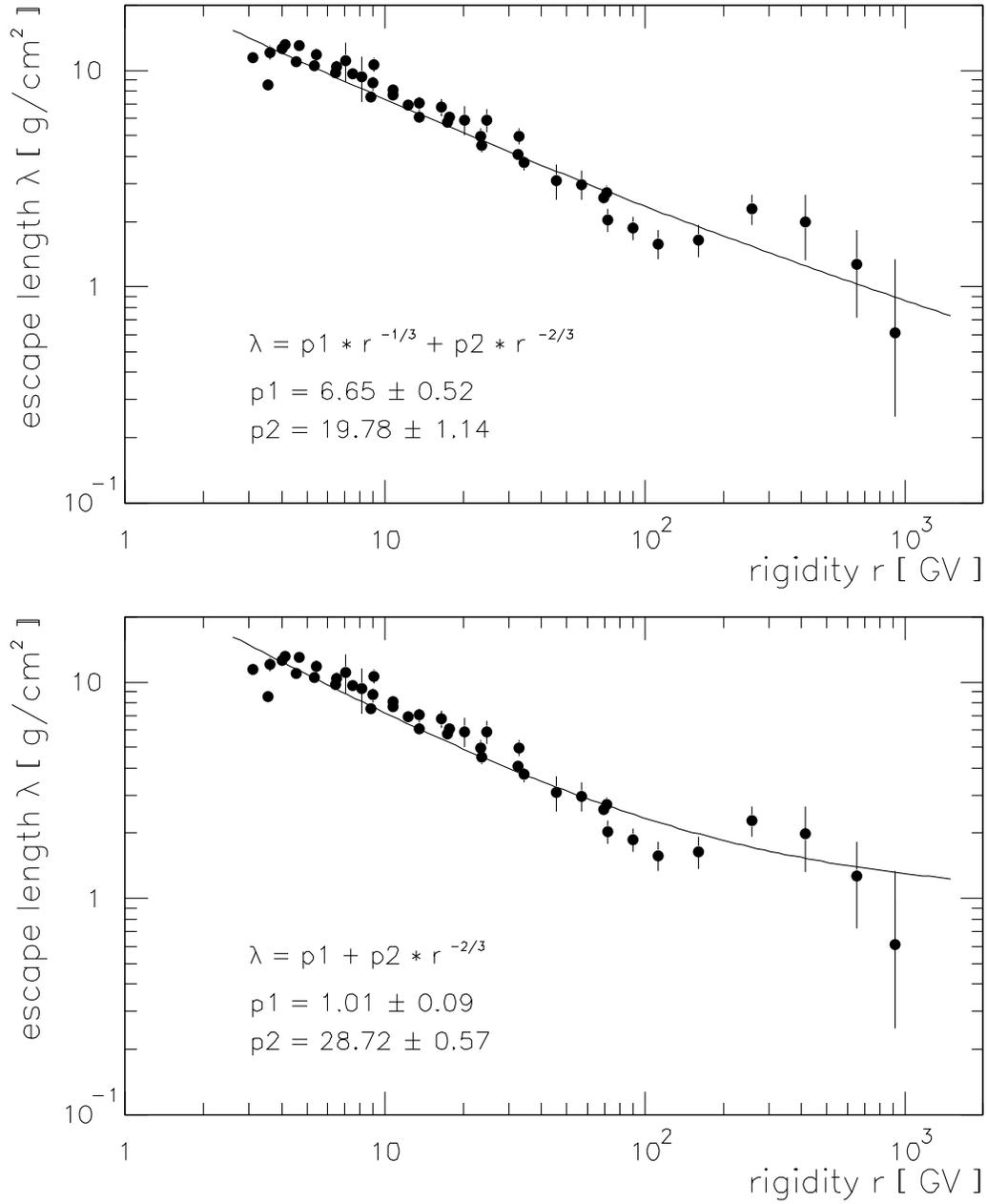
\figure{2}{Here we show the data as given by Shibata (1995), and
fit them with
the two models explained.  The top shows the combination of
process (a) and (b),
and the bottom shows the combination of processes (a) and (c). (For
explanation see text, Sect. 4)}
\endfig

We note that the ratio of these processes depends on the spallation cross
section of the element considered; the relative strength of
process (c) versus
process (b) clearly depends on the element.  For primary elements whose
spallation cross section corresponds approximately to the
effective grammage of
process (c), the spallation of such an element is strongly affected by this
process, while the spallation in the much higher grammage
of processes (a)
and (b) would then proceed to also influence the first generation secondaries,
so as to finally produce secondaries of many generations
down the element sequence.

Translating this result into the language common in the literature, this means
that escape length as measured in gm/cm$^2$ and escape time can no longer used
synonymously.  The escape time is proportional
to $E^{-1/3}$ in the relativistic range of particle energies.  The escape
length as a means to describe interaction has three different regimes, and the
one relevant in the GeV/nucleon range is about $E^{-0.6}$, and
here, in our simplistic model, $\sim E^{-2/3}$.

All three contributions (a), (b), and (c) can be looked for in the data; 
the data are normally shown as {\it escape length}, which basically is the
energy dependence
of the secondary to primary ratio in the steady state leaky box model (see,
e.g., Fig. 20 in Shibata 1995).  With our model we can plot the same data as
Shibata (1995) and can check whether we can fit either process (a)
above combined with
process (b), or process (c).  We do this in Fig. 2.

In the first model combination (process (a) and (b)) we can
satisfactorily fit the
data with a grammage of about 19.8 gm/cm$^2$ for process (a)
and a grammage of
about 6.6 gm/cm$^2$ for process (b); the transition rigidity would be at
27 $\pm$ 5 GV.

In the second model combination (process (a) and (c)) we can
also satisfactorily fit
the data with a grammage of about 29 gm/cm$^2$ for process (a),
rather close to
canonical values, and a grammage of $\approx 1 \, \rm gm/cm^2$ for process (c).

A judgement which model combination is a better to the data
overall will be
made below.  The result is that model combination (a) and (c)
appears to match the
high energy data better for secondary elements; this means
that spallation in
time-dependent molecular clouds and in the molecular shells
around massive star
winds are the dominant contributors to the spallation observed.

We note again, as already emphasized in paper CR I 
that we
use a turbulence spectrum in the interstellar medium, which
has a single
powerlaw over the entire range of length scales relevant
for cosmic ray
scattering, corresponding to energies up to a few EeV, and
have argued that
such a powerlaw is best approximated by a Kolmogorov law
(Wiebel-Sooth et al. (1995); Wiebel-Sooth et al. (paper CR VI),
in prep.; Biermann (1997b)).

\titleb{The elements Li, Be and B}

The nuclei lithium, beryllium, and boron in energetic cosmic
rays are produced
mostly from the breakup of carbon and oxygen nuclei.  The
combined spectrum is
shown in Fig. 3.

\begfig 11 cm
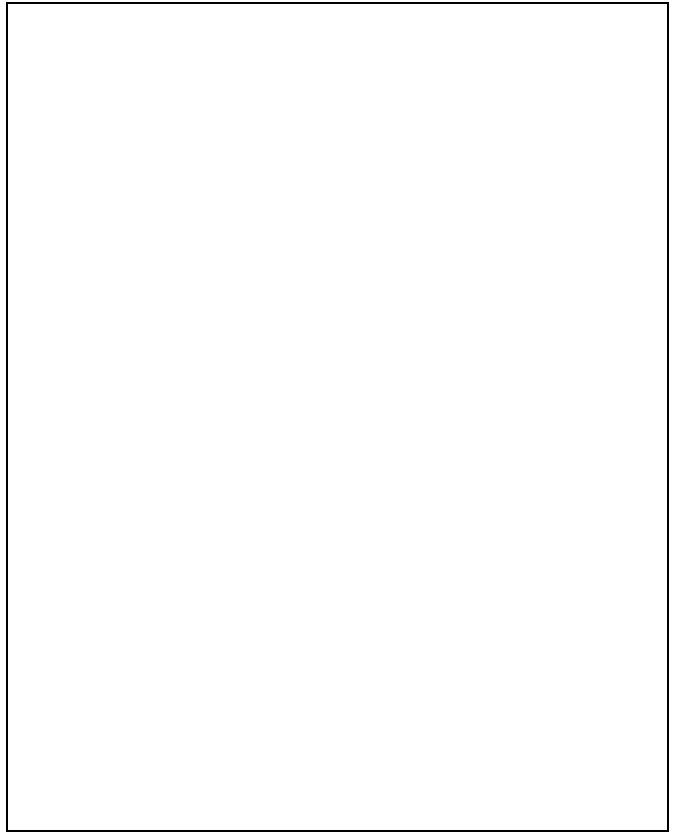
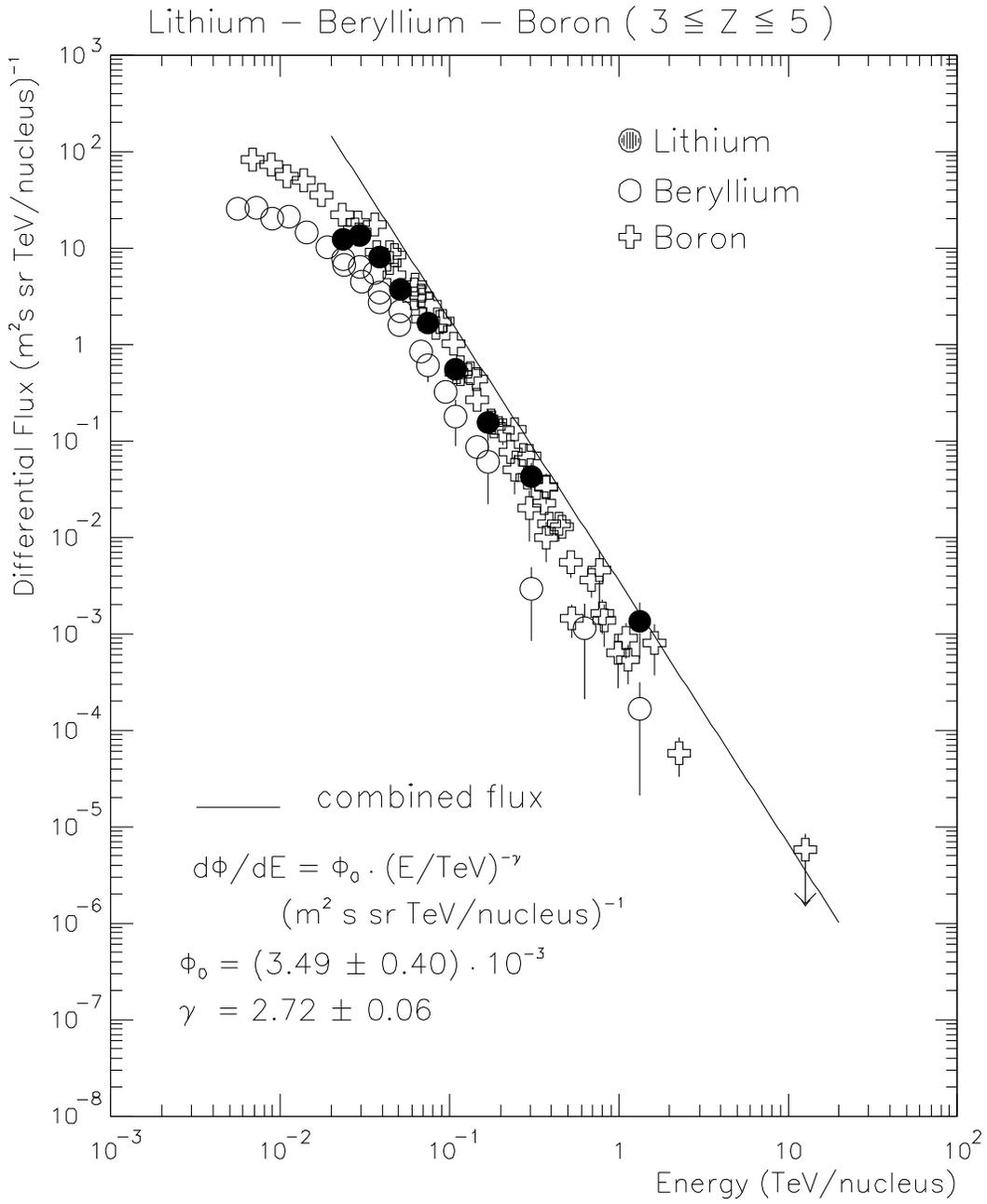
\figure{3}{The data of the differential fluxes for the
elements Li, Be, and B
including the fit for the combined flux.}
\endfig

We note that the spectrum has a fairly large error range, but
is quite
consistent with a source-related component.  However, the
individual spectrum
of Boron suggests a steepening by $\approx \, 1/3$, and so
may be consistent with
process (b), the spallation in clouds at energies which are
so high as to render
the time evolution of clouds irrelevant.  On the other hand,
the error bars for
all three elements are so large, that a certain conclusion
cannot be drawn,
other than that processes (b) and (c) are both compatible
with these high energy
data, and process (a) is hard to reconcile with the data.

\titleb{The sub-Fe group}

The elements scandium through manganese, Sc, Ti, V, Cr and Mn
are mostly
produced in the breakup of the Fe-group elements Fe, Co and Ni.
Their combined
spectrum is shown in Fig. 4.  Here we also show the spectrum,
which would
result from spallation in the interstellar medium at large,
using the notion
that for these high energy particles the time dependence of
molecular clouds is
no longer relevant (process (b)), but still dominant over
source-related
spallation (process (c)). 

It seems that process (c),
the source related spallation gives a much better fit to the data.
But taken into account the statistical errors and systematic uncertainties 
involved in the data measurements, a final decision between the different models
cannot be made at the moment.
Within the errors, both model combinations are compatible with the data.

The spectral index for the sub-Fe
elements is the same as that for the Fe-group elements, consistent with the
notion that at these particle energies the spallation is dominated by source
interaction.

The explanation put forward by Nath \& Biermann (1994b) for the
COMPTEL-observation of $\gamma$-ray emission lines from the Orion star forming
region suggests that the interaction of the supernova shock running at first
through the wind, and then hitting a shell of dense material provides a certain
amount of near-source spallation of a grammage of order $\approx 1 \, \rm
gm/cm^2$.  Here we wish to estimate this grammage from the data to check for
consistency.  Since the charge $Z$ for sub-Fe and Fe-group elements are very
close, it is not important whether we use in this estimate energy/charge or
energy/particle; we will continue to use the latter framework.

The spallation cross section of Fe to sub-Fe is about 300 mbarn,
which corresponds to a grammage of $5.6 \, \rm gm/cm^2$, and so the ratio of
$\Phi_o$ for sub-Fe to Fe-group elements gives us $\approx 0.19$ in numbers
for an inferred grammage of order 1 gm/cm$^2$, quite consistent with the
earlier estimate.  When deriving the spallation during the transport this
source-related grammage has to be subtracted.

\begfig 11cm
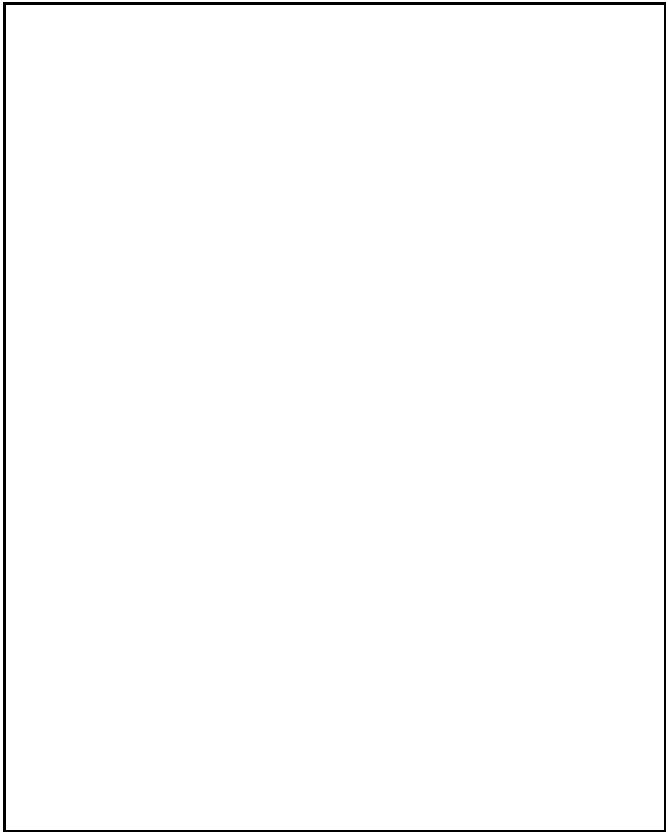
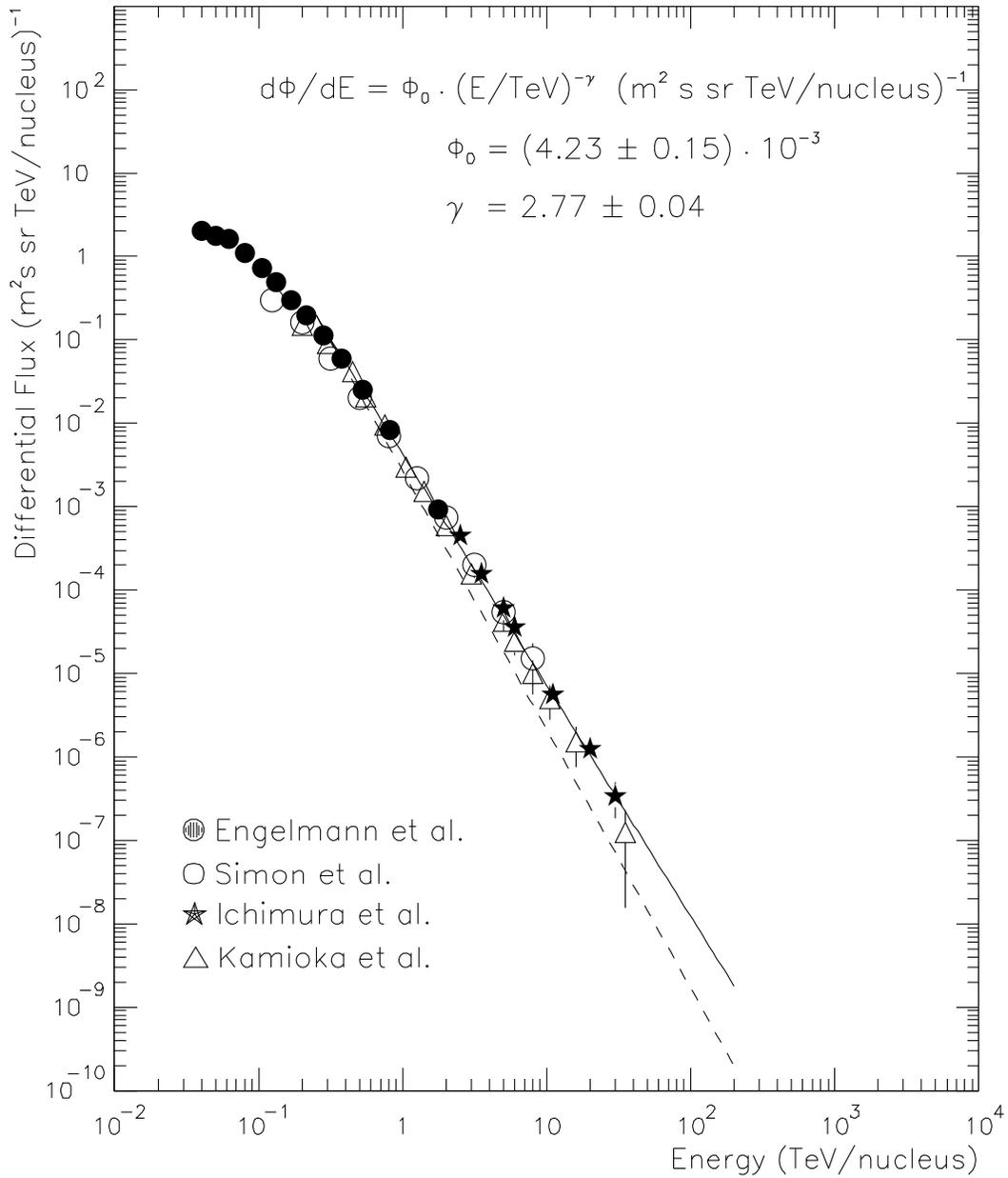
\figure{4}{The spectrum of the sub-Fe group elements : data,
fit (solid line),pure `wind-SN spectrum'' (upper dashed line)
and the ``wind-SN spectrum'' steepened by 1/3 (lower dashed line).
The latter spectra are normalized at that energy where our fit begins,
about 200 GeV/particle. }
\endfig

\titlea {Summary}

Here we have discussed first the predictions and the errors inherent in the
theory to explain the origin of cosmic rays proposed by (paper CR I and later
papers) and then give the cosmic ray spectra above $10 Z$ GeV for all elements
from $Z = 1$ to $28$, based on the full available set of data.  Fitting
powerlaws to the spectra we compare the powerlaw indices with the prediction.

The agreement is quite satisfactory.

The spectra imply that a substantial component of spallation exists at the
source, with a grammage of order $1 \, \rm gm/cm^2$; such a level of
spallation can be understood from the interaction of the cosmic rays in the
supernova shock as it hits a surrounding molecular shell (see Nath \& Biermann
1994b).

Further work will concentrate on improving our understanding of the detailed
abundances of cosmic rays.

\ack{High Energy Physics with author PLB is supported by a NATO travel grant
(CRG 9100072). B. Wiebel-Sooth and H. Meyer are supported by
the BMBF, FRG, under contract number $05\,2\,{\rm WT} \,164$.  PLB would like
to thank Drs. G. Battistoni, T. Gaisser, P. Lipari, D. M{\"u}ller, R. Protheroe,
M. Shapiro, and T. Stanev for many hours of intense discussion  of the topics
raised in this paper.}

\app{}
\vbox{
\centerline{\bf Table 4} \vskip3pt
\vbox{\vfil\hrule
                     \hbox to \hsize{\vrule\hfill
                                     \vbox{\kern10pt
{\settabs \+ Element & References References References\cr
\+ Element & References \cr
\+ H   &  Ryan et al. (1972), Ichimura et al. (1993a, b, c), \cr 
\+     &  Kawamura et al. (1990), Papini et al. (1993), \cr
\+     &  Zatsepin et al. (1990), \cr 
\+     &  Ivanenko et al. (1987, 1990, 1993), \cr 
\+     &  Burnett et al. (1990a, b), Asakimori et al. (1991), \cr
\+     &  JACEE (1993), Shibata (1995) \cr
\+ He  &  Ryan et al. (1972), Ichimura et al. (1993a, b, c), \cr 
\+     &  Kawamura et al. (1990), Papini et al. (1993), \cr
\+     &  Buckley et al. (1993), Dwyer et al. (1993), \cr 
\+     &  Ivanenko et al. (1987, 1990, 1993), \cr 
\+     &  Burnett et al. (1990a, b), Asakimori et al. (1991), \cr
\+     &  JACEE (1993), Shibata (1995) \cr
\+ Li  &  Orth et al. (1978) \cr
\+ Be  &  Engelmann et al. (1985, 1990), Orth et al. (1978), \cr
\+     &  Buckley et al. (1993), Dwyer et al. (1993),\cr
\+ B   &  Engelmann et al. (1985, 1990), Orth et al. (1978)\cr 
\+     &  J\'uliusson (1974), Caldwell (1977),\cr
\+     &  Buckley et al. (1993), Dwyer et al. (1993), \cr
\+     &  Lezniak \& Webber (1978), Webber (1982), \cr
\+     &  Simon et al. (1980), Swordy et al. (1990, 1993), \cr 
\+     &  M\"uller et al. (1991a, b), Grunsfeld et al. (1988) \cr
\+ C   &  Engelmann et al. (1985, 1990), Orth et al. (1978), \cr
\+     &  J\'uliusson (1974), Caldwell (1977), \cr
\+     &  Buckley et al. (1993), Dwyer et al. (1993), \cr
\+     &  Lezniak \& Webber (1978), Webber (1982), \cr
\+     &  Simon et al. (1980), Swordy et al. (1990, 1993), \cr 
\+     &  M\"uller et al. (1991a, b), Grunsfeld et al. (1988) \cr
\+ N   &  Engelmann et al. (1985, 1990), Orth et al. (1978), \cr
\+     &  J\'uliusson (1974), Caldwell (1977),\cr
\+     &  Lezniak \& Webber (1978), Webber (1982), \cr
\+     &  Simon et al. (1980), Swordy et al. (1990, 1993), \cr 
\+     &  M\"uller et al. (1991a, b), Grunsfeld et al. (1988) \cr
\+ O   &  Engelmann et al. (1985, 1990), Orth et al. (1978), \cr
\+     &  Lezniak \& Webber (1978), Webber (1982), \cr 
\+     &  J\'uliusson (1974), Buckley et al. (1993), \cr
\+     &  Dwyer et al. (1993), Caldwell (1977), \cr
\+     &  Simon et al. (1980), Swordy et al. (1990, 1993), \cr 
\+     &  M\"uller et al. (1991a, b), Grunsfeld et al. (1988) \cr 
\+ F   &  Engelmann et al. (1985, 1990), Orth et al. (1978), \cr
\+     &  J\'uliusson (1974), Caldwell (1977) \cr
\+ Ne  &  Engelmann et al. (1985, 1990), Orth et al. (1978), \cr
\+     &  J\'uliusson (1974), Caldwell (1977),\cr
\+     &  Simon et al. (1980), Swordy et al. (1990, 1993), \cr 
\+     &  M\"uller et al. (1991a, b), Grunsfeld et al. (1988) \cr 
\+ Na  &  Engelmann et al. (1985, 1990), Orth et al. (1978), \cr
\+     &  Caldwell (1977) \cr
}\kern10pt}
                                     \hfill\vrule}
                     \hrule\vfil}
}

\vbox{
\centerline{\bf Table 4 (continued) } \vskip3pt
\vbox{\vfil\hrule
                     \hbox to \hsize{\vrule\hfill
                                     \vbox{\kern10pt
{\settabs \+ Element & References References References\cr
\+ Element & References \cr
\+ Mg  &  Engelmann et al. (1985, 1990), Orth et al. (1978), \cr
\+     &  J\'uliusson (1974), Caldwell (1977), \cr
\+     &  Simon et al. (1980), Swordy et al. (1990, 1993), \cr
\+     &  M\"uller et al. (1991a, b), Grunsfeld et al. (1988), \cr 
\+ Al  &  Engelmann et al. (1985, 1990), Orth et al. (1978), \cr
\+     &  J\'uliusson (1974), Caldwell (1977), \cr
\+ Si  &  Engelmann et al. (1985, 1990), Orth et al. (1978), \cr 
\+     &  J\'uliusson (1974), Caldwell (1977), \cr
\+     &  Ichimura et al. (1993a, b, c), \cr 
\+     &  Kawamura et al. (1990), Kamioka et al. (1997), \cr
\+     &  Swordy et al. (1990, 1993), \cr 
\+     &  M\"uller et al. (1991a, b), Grunsfeld et al. (1988) \cr 
\+ P   &  Engelmann et al. (1985, 1990) \cr
\+ S   &  Ichimura et al. (1993a, b, c), \cr 
\+     &  Kawamura et al. (1990), Kamioka et al. (1997), \cr 
\+     &  Engelmann et al. (1985, 1990) \cr
\+ Cl  &  Engelmann et al. (1985, 1990) \cr
\+ Ar  &  Engelmann et al. (1985, 1990), \cr 
\+     &  Kamioka et al. (1997) \cr
\+ K   &  Engelmann et al. (1985, 1990) \cr
\+ Ca  &  Engelmann et al. (1985, 1990),\cr 
\+     &  Ichimura et al. (1993a, b, c), \cr 
\+     &  Kawamura et al. (1990), Kamioka et al. (1997) \cr
\+ Sc  &  Engelmann et al. (1985, 1990) \cr
\+ Ti  &  Engelmann et al. (1985, 1990) \cr
\+ V   &  Engelmann et al. (1985, 1990) \cr
\+ Cr  &  Engelmann et al. (1985, 1990) \cr
\+ Mn  &  Engelmann et al. (1985, 1990) \cr
\+ Fe  &  Engelmann et al. (1985, 1990), Minagawa (1981) \cr
\+ Co  &  Engelmann et al. (1985, 1990) \cr
\+ Ni  &  Engelmann et al. (1985, 1990), Minagawa (1981)\cr
\+ C-O &  Burnett et al. (1990a, b), \cr
\+     &  Asakimori et al. (1991), JACEE (1993) \cr
\+ Cl - Ca & Simon et al. (1980), Burnett et al. (1990a, b), \cr 
\+         & Asakimori et al. (1991), JACEE (1993), \cr 
\+         & Shibata (1995), Ichimura et al. (1993a, b, c), \cr 
\+         & Kawamura et al. (1990), Kamioka et al. (1997) \cr
\+ Sc - Mn & Simon et al. (1980), Ichimura et al. (1993a, b, c), \cr 
\+         & Kawamura et al. (1990), Kamioka et al. (1997) \cr
\+ Fe - Ni & Engelmann et al. (1985, 1990), \cr 
\+         & Simon et al. (1980), Swordy et al. (1990, 1993), \cr
\+         & M\"uller et al. (1991a, b), Grunsfeld et al. (1988), \cr 
\+         & Ichimura et al. (1993a, b, c), \cr
\+         & Kawamura et al. (1990), Kamioka et al. (1997), \cr 
\+         & Burnett et al. (1990a, b), Asakimori et al. (1991), \cr
\+         & JACEE (1993), Shibata (1995) \cr
}\kern10pt}
                                     \hfill\vrule}
                     \hrule\vfil}
}

\begref

\ref Amenomori M., Cao Z., Dai B.Z., et al., 1996, ApJ 461, 408 

\ref  Asakimori K., Burnett T.H., Cherry M.L., et al.,  1991, 
Proc. 22nd ICRC Dublin, OG 6.1-2 and OG 6.2-9

\ref Berezinsky V.S., Bulanov S.V., Dogiel V.A., Ptuskin V.S., 1990,
{\it Astrophysics of Cosmic Rays},  North-Holland, Amsterdam (especially chapter IV)

\ref  Biermann P.L.,  1993a, A\&A  271, 649
(paper CR I)

\ref  Biermann P.L.,  1993b, in ``Invited and rapporteur lectures of the
23rd International Cosmic Ray Conference", Calgary, Eds. R.B. Hicks, and D.
Leahy, World Scientific, Singapore, p. 45 - 83

\ref  Biermann P.L.,  1993c, ``Highly luminous radiogalaxies as sources of
cosmic rays" in ``Currents in Astrophysics and Cosmology", (Conf. proc.
1990, paper sent in 1989) Eds. G.G. Fazio, R. Silberberg, Cambridge Univ.
Press, Cambridge, UK, p. 12 - 19

\ref  Biermann P.L.,  1994, in ``High energy astrophysics", Ed. J. Matthews,
World Scientific, Singapore, p. 217 - 286

\ref Biermann P.L.,  1995a, in ``Frontier objects in astrophysics and particle
physics", Eds. F. Giovanelli et al., Italian Physical Society, Bologna,
p. 469 - 481

\ref Biermann P.L.,  1995b, in ``Proc. of the Gamow seminar in St. Petersburg",
Eds. A.M. Bykov, R.A. Chevalier, Spac. Sci. Rev. 74, 385 - 396

\ref Biermann P.L.,  1996, in Proc. conference MESON96, Krakow, Eds. A. Magiera
et al., Acta Phys. Polon. B 27, 3399 - 3415

\ref Biermann P.L.,  1997a, J. Phys. G 23, 1

\ref Biermann P.L.,  1997b, in ``Cosmic winds and the heliosphere", Eds. J.R.
Jokipii et al., Univ. of Arizona Press, (76 pages) (in press) 

\ref  Biermann P.L., Cassinelli J.P.,  1993 , A\&A  277, 691
(paper CR II)

\ref  Biermann P.L., Strom, R.G., 1993, A\&A 275, 659
(paper CR III)

\ref Biermann P.L., Gaisser T.K., Stanev T.,  1995a, Phys. Rev. D 51, 3450

\ref Biermann P.L., Strom R.G., Falcke H.,  1995b, A\&A 302, 429 (paper CR V)

\ref Biermann P.L., Kang H., Ryu D.,  1997, in ``Extremely high energy cosmic
rays", Ed. M. Nagano, Univ. of Tokyo, (in press) 

\ref  Buckley J., Dwyer J., M\"uller D., et al., 
1993, Proc. 23rd ICRC Calgary, vol. 1, 599 - 602

\ref  Burnett T.H., Dake S., Derrickson J.H., et al.,  1990a, ApJ 349, L25

\ref  Burnett T.H., Dake S., Derrickson J.H., et al.,  
1990b, Proc. 21st ICRC Adelaide, OG 6.1-10

\ref  Caldwell J.H.,  1977, ApJ 218, 269

\ref  Drury L.O'C,  1983, Rep. Prog. Phys. 46, 973

\ref  Drury L.O'C, Aharonian F.A., V\"olk H.J.,  1993, Proc. 23rd ICRC Calgary,
vol. 1, 455 - 458

\ref DuVernois M.A., Garcia-Munoz M., Pyle K.R., et al.,  1996a, ApJ 466, 457

\ref DuVernois M.A., Simpson, J.A., Thayer, M.R.,  1996b, A\&A 316, 555

\ref  Dwyer J., Buckley J., M\"uller D., et al., 
1993, Proc. 23rd ICRC Calgary, vol. 1 , 587 - 590

\ref  Engelmann J.J., Ferrando P., Soutoul A., et al.,  1990, A\&A 233, 96

\ref  Engelmann J.J., Goret P., J\'uliusson E., et al.,  1985, A\&A 148, 12

\ref  Evenson P., Meyer P.,  1984, J. Geophys. Res. 89, No. A5, 2647

\ref  Fermi E.,  1949,  Phys. Rev. 2nd ser., vol. 75, no. 8, 1169

\ref  Fermi E.,  1954, ApJ. 119, 1

\ref  Garcia-Munoz M., Meyer P., Pyle K.R., Simpson J.A.,  
1986, J. Geophys. Res. 91, No. A3, 2858

\ref Garcia-Munoz M., Simpson J.A., Guzik T.G., et al.,  1987, ApJS 64, 269 

\ref Goldstein M.L., Roberts D.A., Matthaeus W.H.,  1995, ARA\&A 33, 283

\ref  Grunsfeld J.M., L'Heureux J., Meyer P., et al.,  1988, ApJ 327, L31

\ref Hess V.F., 1912, Phys. Z. 13, 1084

\ref  Ichimura M., Kamioka E., Kirii K., et al., 1993a, Proc. 23rd ICRC Calgary,
vol. 2, 1 - 4, vol. 2, 5 - 8 and vol. 2, 9 - 12

\ref  Ichimura M., Kogawa M., Kuramata S., et al., 1993b, Phys. Rev. D, vol.48, No.5, 1949

\ref  Ichimura M., Kogawa M., Kuramata S., et al., 1993c, 
Proc. 23rd ICRC Calgary, vol. 1, 591 - 594

\ref  Ivanenko I.P., Grigorov N.L., Shestoperov V.Ya., et al.,  
1987, Sov. J. Nucl. Phys. 45 (4), 664

\ref  Ivanenko I.P., Rapoport I.D., Shestoperov V.Ya., et al.,  
1990, Proc. 21st ICRC Adelaide, OG 6.1-3

\ref  Ivanenko I.P., Shestoperov V.Ya., Chicova L.O.,  et al.,  
1993, Proc. 23rd ICRC Calgary, vol. 2, 17 - 20

\ref  JACEE-Collaboration,  1993, Contributions to 23rd ICRC
         Calgary, Louisiana State University Report, LA 70803 - 4001

\ref Jokipii J.R.,  1982, ApJ 255, 716

\ref  Jokipii J.R.,  1987, ApJ 313, 842

\ref  Jokipii J.R., Levy E.H., Hubbard W.B., 1977,  ApJ 213, 861

\ref  J\'uliusson E.,  1974, ApJ, 191, 331 

\ref Kamioka E., Hareyama M., Ichimura M., et al., 1997, Astropart. Phys. 6, 155

\ref  Kawamura Y., Matsutani H., Nanjyo H., et al., 
1990, Proc. 21st ICRC Adelaide, OG 6.1-5 and OG 6.1-6

\ref  Kohlh\"orster W., 1913,  Phys. Z. 14, 1153

\ref  Lezniak J.A., Webber W.R.,  1978 , ApJ, 223, 676

\ref  Minagawa G.,  1981, ApJ 248, 847

\ref  M\"uller D., Grunsfeld J., L'Heureux J., et al.,  
1991a, Proc. 22nd ICRC Dublin, OG 6.1.12

\ref  M\"uller D., Swordy S.P., Meyer P., et al,  1991b, ApJ 374, 356

\ref Nagano M., Hara T., Hatano Y., et al.,  1984 J. Phys. G 10, 1295 

\ref Nath B.B., Biermann P.L.,  1993,  MNRAS 265, 241

\ref Nath B.B., Biermann P.L.,  1994a,  MNRAS 267, 447

\ref Nath B.B., Biermann P.L.,  1994b,  MNRAS 270, L33

\ref  Orth C.D., Buffington A., Smoot G.F., et al., 1978, ApJ 226, 1147

\ref Papini P., Grimani C., Basini F., et al.,  1993, Proc. 23rd ICRC Calgary, vol. 1, 579

\ref  Parker E.N.,  1958,  ApJ. 128, 664 

\ref Peters B.,  1959,  Nuov. Cim. Suppl. vol. XIV, ser. X, 436

\ref Peters B.,  1961,  Nuov. Cim. vol. XXII, 800 

\ref  Prandtl L.,  1925, Zeitschrift angew. Math. und Mech. 5, 136

\ref Prandtl L.,  1949, ``Guide through the theory of fluid motion (German)"
Vieweg, Braunschweig, fifth ed. (1st Ed. ``Abri{\ss} der Str\"omungslehre" 1931)

\ref  Rachen J., Biermann P.L.,  1993, A\&A. 272, 161 (paper UHE  CR I)

\ref  Rachen J., Stanev T., Biermann P.L.,  1993, A\&A
273, 377 (paper UHE CR II)

\ref Ramaty R., Kozlovsky B., Lingenfelter R.E., Reeves H., 1997, ApJ 488, 730

\ref Rickett B.J.,  1990, ARA\&A 28, 561 

\ref  Ryan M.J., Ormes J.F., Balasubrahmanyan V.K.,  1972, 
Phys. Rev. Lett. 28, No.15, 985

\ref  Seo E.S., Ormes J.F., Streitmatter R.E.,  et al.,  1991, ApJ 378, 763 

\ref  Simon M., Spiegelhauer H., Schmidt W.K.H., et al.,  1980, ApJ 239, 712

\ref Shibata T.,  1995, Rapporteur paper in ``Proceedings of the 24th 
International Cosmic Ray Conference Rome, Invited, Rapporteurs \& Highlight
Papers", Eds. N. Iucci, E. Lamanna, pp. 713-736 

\ref  Stanev T., Biermann P.L., Gaisser T.K.,  1993 , A\&A
274, 902 (paper CR IV)

\ref Stanev T., Biermann P.L., Lloyd-Evans J., Rachen J., Watson A.A.,
1995, Phys. Rev. Lett. 75, 3056

\ref Str{\"o}mgren B., 1953, in ``The Sun", Ed. G.P. Kuiper, Univ. of Chicago
Press, p. 36 - 87

\ref  Swordy S.P., L'Heureux J., Meyer P., et al.,  1993, ApJ 403, 658 

\ref  Swordy S.P., M\"uller D., Meyer P., et al.,  1990, ApJ 349, 625

\ref  Webber W.R.,  1982, ApJ, 255, 329

\ref Wiebel-Sooth B., Biermann P.L., Meyer H., 1995, Proc. 24th ICRC Rome,
vol. 3, 45 - 47

\ref Wiebel-Sooth B., Biermann P.L., Meyer H., (paper CR VI), in prep.

\ref  Zatsepin V.I., Zamchalova E.A., Varkovitskaya A.Ya., et al., 
1990, Proc. 21st ICRC Adelaide, OG 6.1-4

\endref

\bye